%% file: edgepaper.tex
\documentclass[twocolumn,twocolappendix,astrosym]{openjournal}

\usepackage{graphicx}

\usepackage{latexsym,amssymb}

\usepackage{amsmath,morefloats}
\usepackage[backref,breaklinks,colorlinks,citecolor=blue]{hyperref}
\usepackage{natbib,subfigure,color}
\usepackage[usenames,dvipsnames]{xcolor}
\usepackage{verbatim}
\usepackage{threeparttable}

\usepackage{pgfplots}
\pgfplotsset{compat=1.18}
\usepgfplotslibrary{fillbetween}
\usepgfplotslibrary{groupplots}
\usetikzlibrary{patterns}

\input{commands.tex}

\input{colors.tex}

\shorttitle{Edges in Coadd Images}
\shortauthors{Sheldon}

\begin{document}

\title{Edges in Coadded Images}
\author{Erin S. Sheldon}
\affil{Brookhaven National Laboratory, Bldg 510, Upton, NY 11973, USA}



\begin{abstract}

    We investigate how discontinuities in the point spread function (PSF) and
    image noise affect weak gravitational lensing shear measurements. Our focus is on
    discontinuities that arise in coadded images, particularly when edges from
    input images cross the coadd region. Using \mcal\ and \mdet\ for shear
    calibration and a simple weighted mean coaddition, we find no significant
    biases for typical galaxy populations, typical edge hit rates (a few
    percent), or coadds with tens to hundreds of input images. Biases exceeding
    requirements for large lensing surveys occur only in extreme conditions:
    (a) coadds with just two input images; (b) an image edge crosses the object
    in about 25\% of coadds; (c) PSF size variations greater than 25\%; (d)
    relatively small galaxies.  Even in these extreme cases, accurate shear
    recovery is achievable by identifying and removing problematic measurements.
    We use a simple statistic that quantifies the relative
    variation in PSF size across each object.

\end{abstract}

\section{Introduction} \label{sec:intro}


Gravitational lensing caused by cosmic structures distorts and shears galaxy
images. By measuring these distortions, we gain unique insights into the
universe's structure and evolution \citep[for a recent review,
see][]{PratWLReview2025}.  Shear measurements have emerged as an important tool
for investigating phenomena like dark matter and the universe's accelerated
expansion. Consequently, accurate shear estimation is a central objective of
current and forthcoming imaging surveys such as the Rubin Observatory Legacy
Survey of Space and Time
\citep[LSST,][]{IvezicLSST2008,MandSysReview2018,LSSTScienceBook2009}, the
Euclid Mission \citep{EuclidMission2025} and the Roman Space Telescope Survey
\citep{TroxelRoman2021}.

To achieve accurate shear measurements, it is essential to thoroughly
characterize the images used for analysis, including background noise and the
point spread function (PSF) of the atmosphere, telescope and detectors.  In
this paper we investigate the impact of discontinuous PSF and noise resulting
from edges in coadded images, and their implications for shear estimation.

In the weak lensing approximation, shear can be inferred from observables such
as galaxy ellipticity.  However, ellipticity measurements are performed on
noisy, PSF-convolved images (where the PSF acts locally and independently of
the shear), leading to biased shear estimates \citep{MandSysReview2018}.
Contemporary techniques, including \mdet\ \citep{mdet2020,mdetlsst2023}, Anacal
\citep{AnacalModeling2025}, and BFD \citep{BernBFD2016} can accurately
calibrate ellipticity as a shear proxy, provided accurate models for both the
noise and PSF.

Inaccuracies in the PSF model introduce biases into shear estimates. For illustration,
consider ellipticities derived from second moments $I_{kl}$ of the light profile $I$:
\begin{align} \label{eq:moments}
    I_{kl} &= \frac{\sum_{i,j} I(i, j) (x-i)^k (y-j)^l}{\sum_{i,j} I(i, j)} \nonumber \\
    T & \equiv I_{11} + I_{22} \nonumber \\
    e_1 &= \frac{(I_{11} - I_{22})}{T} \nonumber \\
    e_2 &= 2 \frac{I_{12}}{T} \nonumber
\end{align}
where $i,j$ run over pixels in the image, and $x, y$ is the center of the object.
Convolution with the PSF adds its moments to those of the galaxy, resulting in
$\tilde{I}_{ij} = I_{ij} + I^{\mathrm{PSF}}_{ij}$.  If the galaxy's raw moments
$\tilde{I}_{ij}$ were known precisely (though noise prevents this), PSF
correction could be achieved by subtracting the PSF moments\footnote{These raw
unweighted moments are impractical for ellipticity measurement on noisy images
due to their low precision.  Instead, model fitting or weighted moments are
employed to improve the precision. Nonetheless, the insights from unweighted
moments remain applicable.}.

Suppose the PSF estimate is biased, with inferred moments
$I_{ij}^{\mathrm{PSF}} + \Delta I_{ij}^{\mathrm{PSF}}$.   Correcting with this
biased PSF yields biased galaxy ellipticities.  For small PSF moment errors,
the inferred ellipticity $\tilde{e}$ is, to leading order:
\begin{equation} \label{eq:ebias}
    \tilde{e}_i \approx e_i \left(1 + \frac{\Delta T^{\mathrm{PSF}}}{T}\right) - \Delta e_i^{\mathrm{PSF}} \frac{\Delta T^{\mathrm{PSF}}}{T},
\end{equation}
where $i \in \{1, 2\}$, and $\Delta e_1^{\mathrm{PSF}} \equiv (\Delta I_{11}^{\mathrm{PSF}} - \Delta I_{22}^{\mathrm{PSF}}) / \Delta T^{\mathrm{PSF}}$ (with a similar expression for $e_2$).
This bias becomes negligible when $\Delta T^{\mathrm{PSF}} \ll T$, 
meaning PSF moment errors are small relative to galaxy moments.

For weak shear measurements obtained by averaging ellipticities, the estimated
mean shear \gest\ can be modeled linearly in terms of the true shear
\vecgam$=(\gamma_1, \gamma_2)$:
\begin{equation} \label{eq:biaslin}
    \gest = \vecgam \times \vecm + \vecc,
\end{equation}
where \vecm\ denotes multiplicative bias and \vecc\ a shear independent bias.
Comparing Eqs. \ref{eq:biaslin} and \ref{eq:ebias} reveals that
PSF errors can contribute to both \vecm\ and \vecc.

This study examines a specific bias source arising from discontinuities in
coadded images. Coaddition involves remapping overlapping images onto a common
pixel grid and averaging to increase the signal-to-noise ratio.  Coadding also
has the benefit of reducing the computational cost to measure objects.  The
original images can be used for shear measurement \citep{GattiY3Shear2021}.
However, scaling up to hundreds of images per object might be computationally
prohibitive.

Furthermore, the detection process itself is shear dependent \citep{mdet2020}
and introduces a selection bias.  The \mdet\ shear measurement method was
developed to calibrate this bias. It works by repeating detection and object
measurement on images that have been artificially sheared.  But detection
algorithms typically work on a single image, not on a set of images, for
reasons of both simplicity and computational efficiency.   Indeed the software
used by the aforementioned surveys will, at least initially, use standard
detection algorithms.  Thus, from a practical standpoint, \mdet-like algorithms
need to work on coadds.

However, if an input image partially covers the coadd region, its
edge traverses the coadd. Variations in PSF among input images lead to
discontinuities in the coadd PSF. A single continuous model cannot fully
describe the PSF of such a coadd, resulting in biased PSF corrections.  In
particular, if the image of an object is crossed by an edge, no single PSF can
be used to correct measurements of the object's ellipticity.

The bias from discontinuities is negligible if the PSF variations between
images are sufficiently small (see Equation \ref{eq:ebias}).  Moreover, with an
infinite number of input images, edges induce no bias provided that images on
both sides of an edge sample the same distribution of image properties.  Both
sides necessarily represent fair samples.   In contrast, small numbers of
inputs with large PSF variations relative to galaxy sizes yield non-zero
biases.

Comprehending coadd accuracy limits is important for shear measurement
precision as well as accuracy.  If edges are not permissible, the offending
images must be discarded.  The fraction of discarded images depends on the
dimensions of the coadd and ``dither pattern'', how telescope pointings are
distributed across the sky \citep{ArmstrongCoadd2024}.  Such ``edgeless''
coadds were employed in the Dark Energy Survey year 6 shear analysis
\citep{DESY6Shear2025}, resulting in approximately \deslost\ of the images being
removed. Allowing edges preserves these images.

Alternatively, the input images could be ``homogenized'' so all PSFs match, but
accompanied by a significant increase in noise
\citep{DESDM2011,ZackayCoadd2017}.  Permitting edges not only avoids increased
noise, but allows for potential optimizations in the coaddition process
\citep{KaiserCoadd2004,ZackayCoadd2017}.

Noise discontinuities will also emerge in coadds from varying input noise
levels in the input images, common in ground- and space-based surveys.  \mdet\
can incorporate non-homogeneous noise into its calibration, which may provide a
mechanism to mitigate the effects of noise discontinuties.

We subsequently investigate these effects through simulations emulating LSST
data.  The paper is laid out as follows: In \S \ref{sec:sim} we describe the
simulations used to test edges.  In \S \ref{sec:methods} we describe our
methods for coaddition, image processing and shear measurement. In \S
\ref{sec:results} we show our results.  In \S \ref{sec:diagnostic} we
discuss how to diagnose problem measurements that may cause a shear bias.  In
\S \ref{sec:discuss} we discuss the implications of our results and in \S
\ref{sec:summary} we provide a summary.

\section{Simulations} \label{sec:sim}

In order to assess the impact of PSF and noise discontinuities on shear
measurements, we employed two distinct types of simulations. The first,
referred to as the realistic simulation, replicates the complete LSST focal
plane comprising 189 CCDs, incorporating semi-realistic point spread functions
(PSFs), galaxies, noise, and random sky pointings. Edges arise naturally in the
coadds due to the pointing pattern, camera rotation, and CCD layout, rendering
these simulations similar to anticipated real data.

The second type, called the simple simulation, features fixed galaxy
morphology, low noise, a simpler PSF model, and edges that were randomly
inserted into the images with specified statistical properties. This simulation
does not reflect real data, but is useful for isolating any sources of bias.

We describe each simulation in more detail below.

\subsection{Realistic Simulations} \label{sec:realsim}

\subsubsection{Features of the Realistic Simulations}

\begin{figure*}[t]
    \centering
    \includegraphics[width=\textwidth]{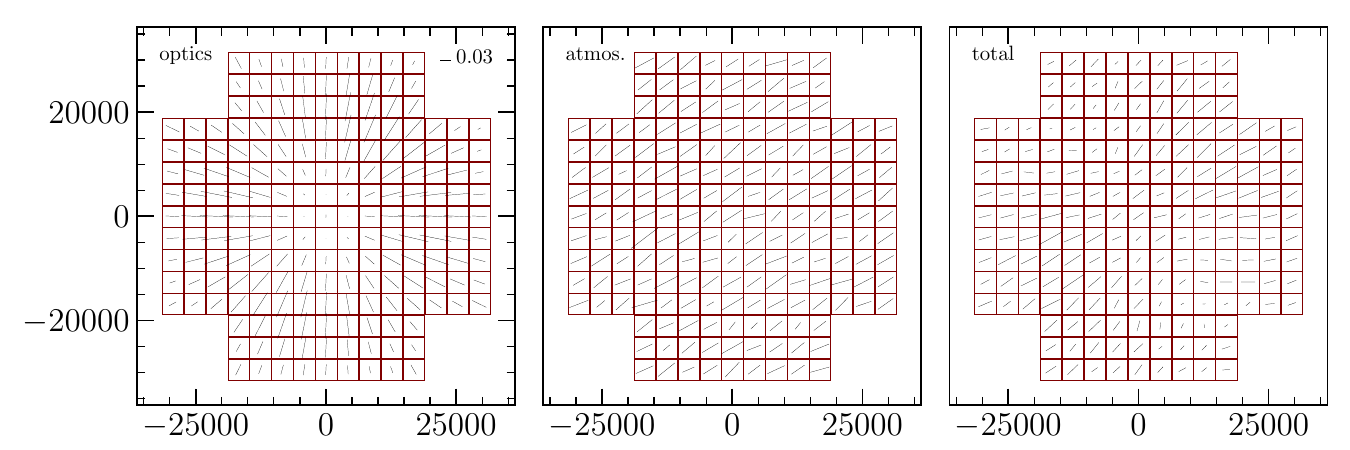}

    \caption{
        Example PSF ellipticity whiskers from the realistic simulation, showing
        the optics (left panel), atmosphere (middle panel), and total combined
        optics, atmosphere, CCD diffusion, and pixelization (right panel).
        Overlaid are lines representing CCD boundaries.  The units of the axes
        are in pixels, and an example whisker length is provided in the left
        panel for scale.
    }

 \label{fig:focalplanewhisker}

\end{figure*}

For the realistic simulation, we modeled the LSST camera as 189 science CCDs,
with 22.2 arcsec gaps between CCDs and a fixed pixel scale of 0.2 arcsec.  We
show the geometry in Figure \ref{fig:focalplanewhisker}.  The PSF in these
simulations was modeled using three distinct, simplified components: an
atmosphere, the optics, and local diffusion effects on each CCD.  

The atmospheric component of the PSF was generated using the power
spectrum-based PSF from the \descwlshearsims\ package, used for the
\cite{mdetlsst2023} lensing simulations.  This PSF was tuned to have mean full
width at half maximum (FWHM) of 0.625 arcsec prior to pixelization.

The optical component drew inspiration from the realistic optical PSF produced
by the \imsim\ package \footnote{\url{https://github.com/LSSTDESC/imSim}}.  In
\imsim\ the optics are fixed for all images and have circular symmetry.  For
this work, the spatial variations were approximated with order-6 Zernike
polynomials \citep{Zernike1934}, constrained to have circular symmetry.  The
profile was modeled as an elliptical Moffat function \citep{Moffat1969}.  The
model was tuned to yield a mean FWHM of approximately 0.35 arcsec before
pixelization.

Additionally, a random Gaussian diffusion kernel was applied to each CCD. This
mimics the lateral spreading of photo-generated charge carriers (electrons)
within the silicon substrate during the charge-collection process.  The FWHM
for each kernel was drawn from a normal distribution having a mean of 0.2
arcsec, matching measurements in LSSTCam \citep{RoodmanCamera2024}, and a
standard deviation $\sigma=0.02$, prior to pixelization.

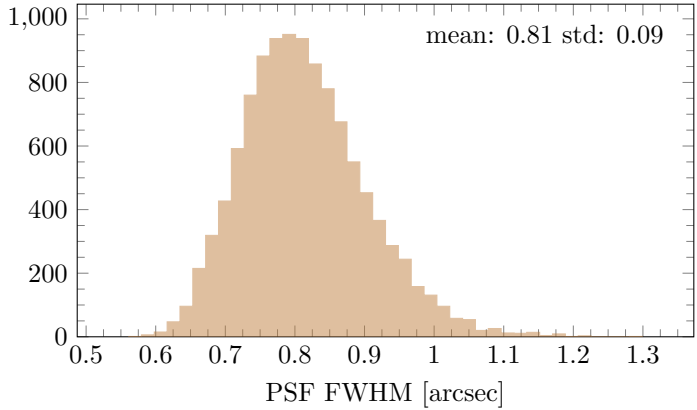
\begin{figure}[h]
    \centering
    \input{figures/psf-fwhm.tex}

    \caption{
        Distribution of PSF FWHM in the realistic simulation.  The FWHM was
        measured on the full PSF, including atmosphere, optics, diffusion,
        and pixelization.
    }

 \label{fig:fwhm}

\end{figure}

These three PSF components were tuned to produce a mean seeing FWHM of about
0.8 arcsec and variations with $\sigma=0.06$ arcsec after pixelization .   The
distribution of PSF FWHM for the full set of simulated exposures is shown in
Figure \ref{fig:fwhm}.  Note that, owing to the non-Gaussian shapes of the
optical and atmospheric profiles, the individual FWHM values quoted above do
not add quadratically to yield 0.8 arcsec. In the regime of large atmospheric
seeing, the optical contribution becomes negligible, whereas under small seeing
conditions, the optical part dominates.

\begin{figure*}[t]
    \centering
    \includegraphics[width=\textwidth]{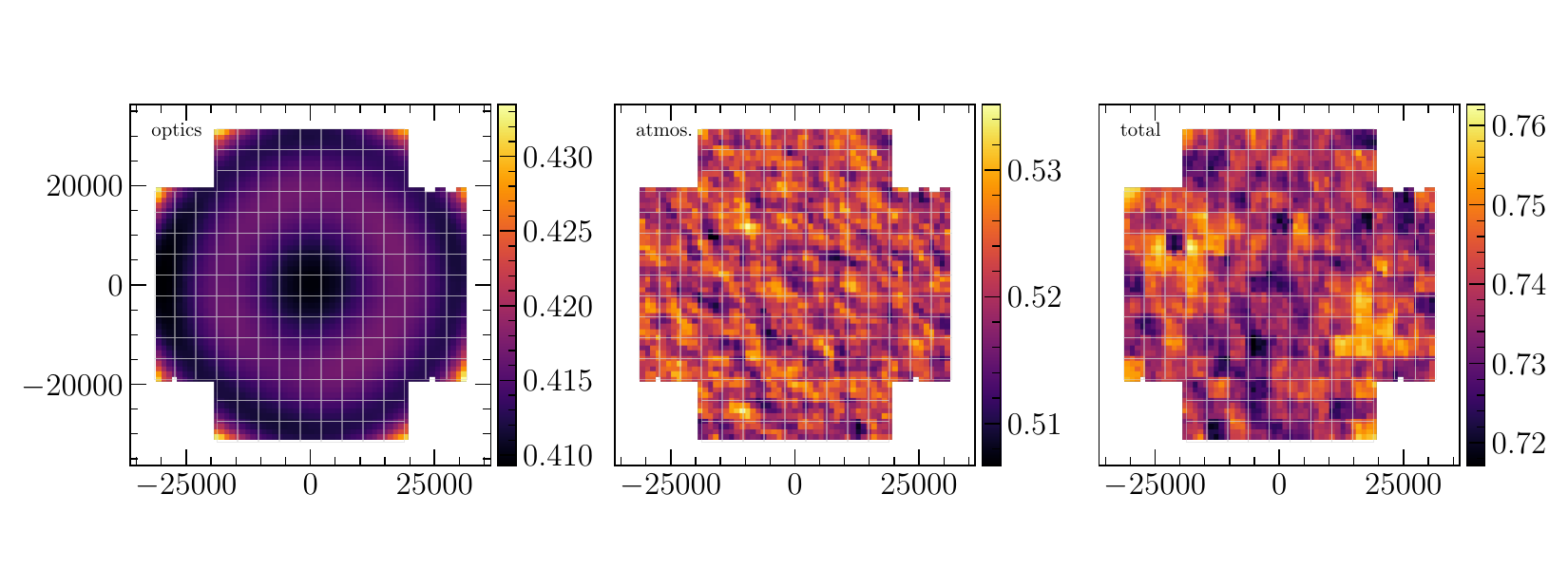}

    \caption{
        Example PSF FWHM in arcsec from the realistic simulation, showing the
        optics (left panel), atmosphere (middle panel), and
        total combined optics, atmosphere, CCD diffusion, and pixelization
        (right panel).  Overlaid are lines representing CCD boundaries.  The
        units of the axes are in pixels.
    }

 \label{fig:focalplanefwhm}

\end{figure*}

An example PSF whisker pattern is displayed in Figure
\ref{fig:focalplanewhisker}, and the corresponding FWHM in Figure
\ref{fig:focalplanefwhm}.

Galaxies were modeled using the profiles provided by the \descwl\ package
\citep{DESCWLSanchez2021}\footnote{\url{https://github.com/LSSTDESC/WeakLensingDeblending}}.
These models are bulge \citep[de Vaucouleurs profile,][]{devauc1948} plus disk
(exponential profile), with the two components having different ellipticities
and scale lengths.  We did not include stars in the simulation.

The LSST imaging survey was simulated by generating random pointings across a
sky region with the expected density based on the operation simulation
database used for the LSST data challenge 2 simulation
\citep[DC2,][]{DC2Abolfathi2021}.  For each simulated pointing, the camera
orientation was randomized to replicate rotational dithers. Note that LSST
pointings and rotational dithers will be forced to be uniform at years 4, 7 and
10, but will not be uniformly random at intermediate stages
\citep{LeistedtUniformRolling2025}.

Sky background levels were modeled to emulate the distribution expected for
LSST $i$-band data, again based on the operation simulation database used for
DC2.

\subsubsection{Images in the Realistic Simulations}

The Year 10 LSST survey is projected to accumulate approximately 450 images per
point in the survey across the combined primary lensing bands $r$, $i$, and $z$
\citep{SRD}.  Generating a complete set of such images, remapping (a.k.a.
warping) them onto a common grid and coadding them would be prohibitively
expensive.

However, several reasonable approximations can accelerate the simulations.
Prior studies indicate that the warping and coadding processes does not produce
any bias for \mcal\ or \mdet\ \citep{ArmstrongCoadd2024,mdetlsst2023}.  Thus,
warping is unnecesary for this test.  As warping is the most time-consuming
aspect of the coadding process, we can save significant computing time by
instead drawing directly into a ``coadd'' image.

Drawing into full images is significantly slower than rendering into smaller
``postage stamps''.  We conducted one computationally intensive simulation with
full images intersected to produce edges, and performed full \mdet\ processing.
This simulation was used to test whether detection effects can be calibrated even in
the presence of edges.  For the other simulations, we bypassed
full images and instead drew directly into stamps.  These stamps were 
intersected with the input image geometry to create edges, replicating
the process used for the full images.  The postage stamp size for each object
was determined by the \galsim\ drawing code.

A uniform shear $\vecgam = (0.02, 0.00)$ was applied to all images.

\subsection{Simple Simulations} \label{sec:simsim}

The simple simulations featured a fixed galaxy profile, variable but spatially
constant PSF, and low noise.  The galaxy was represented as a round exponential
profile with flux chosen to provide high signal-to-noise ratio in the coadd.
The half-light radius was varied to assess how PSF errors propagate into shear
measurements as a function of galaxy size.

The PSF was modeled as an elliptical Moffat profile, with FWHM drawn from a
truncated lognormal distribution (minimum of 0.6 arcsec). The distribution's
width was adjusted as part of the tests. Ellipticities were sampled from a
truncated ($|e| < 1$) isotropic Gaussian in $e_1, e_2$, with $\sigma=0.02$.

As with the realistic simulations, each image was given the same shear $\vecgam
= (0.02, 0.00)$.

\section{Methods} \label{sec:methods}

In this section we present our methods for creating coadd images, performing
image processing and galaxy measurements, and calibrating the shear.

\subsection{Coadding and Edges}

In the realistic simulation, the CCD geometry and sky pointing patterns led to
non-uniform coverage, resulting in natural edges within the coadd images.

For the simple simulation, edges were generated randomly with a specified
probability, enabling experimental control. These edges were created by
randomly selecting a line that passes through the input image, potentially at
an angle to emulate camera rotations.


For all cases we performed a weighted mean coadd.  The weight for each input
image was set to the median of the inverse variance of that image.


When coadding, we did not render and remap actual images. Instead, we rendered
directly in the coadd image coordinates.  To do this we used the RA and DEC of
the object to generate the analytic, pre-pixelization PSF at the corresponding
position in the input image. The galaxy, represented as a \galsim\ object, was
then convolved with this PSF and rendered in coadd coordinates with
pixelization.

\begin{figure}[h]
    \centering
    \includegraphics[width=1.0\columnwidth]{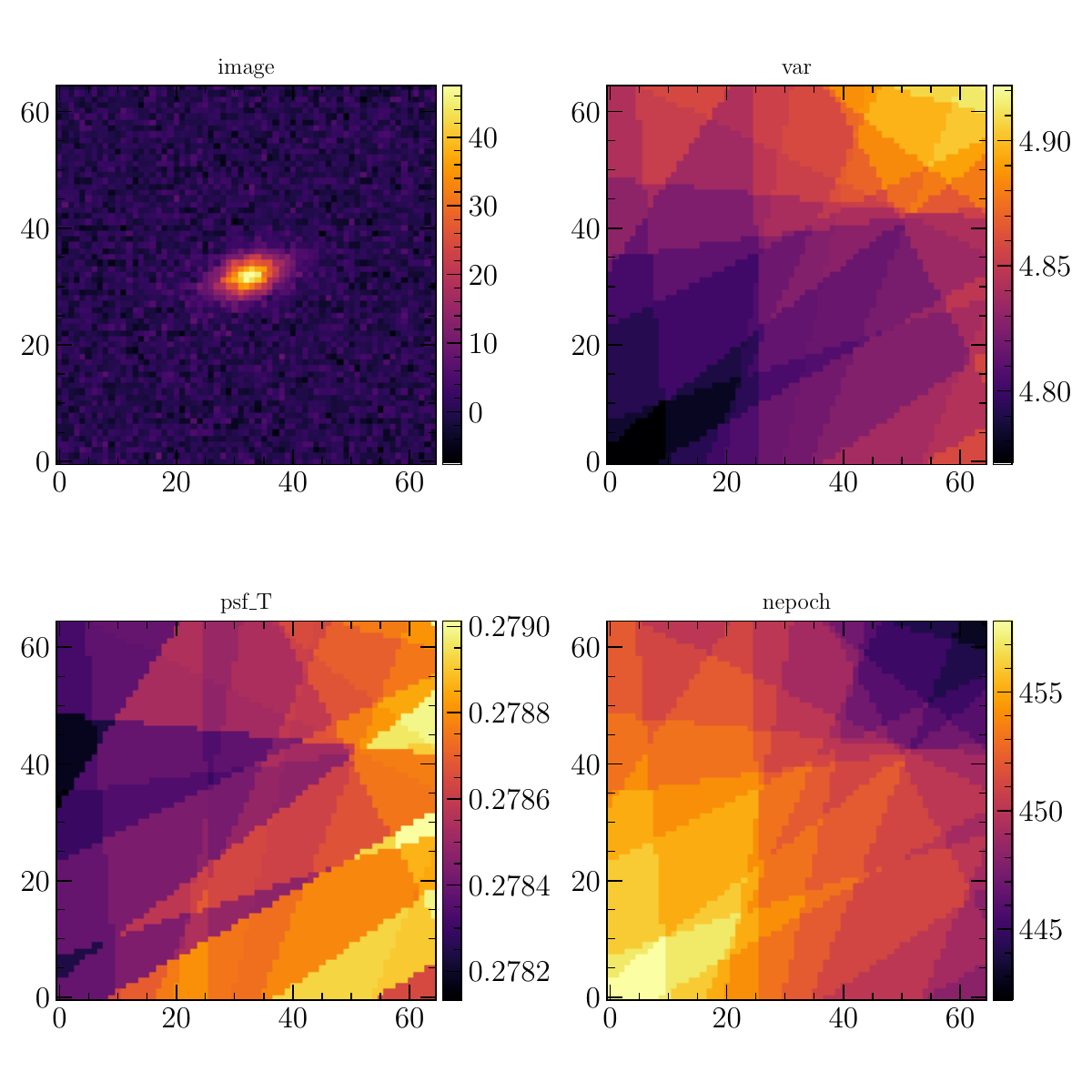}

    \caption{
        Example postage stamps from the realistic simulations.  Upper left: the
        coadded image.  Upper right: map of the variance in each pixel.  Lower
        left: map of the mean \Tpsf\ in each pixel.  Lower right: map of the
        number of images contributing to each pixel.
    }

 \label{fig:stamp}

\end{figure}

This process introduced discontinuities in coadd properties, such as noise and
PSF, at the edges. An example postage stamp from a full-depth realistic
simulation is shown in Figure \ref{fig:stamp}, along with maps
of the noise variance, \Tpsf, and number of epochs contributing to the coadd.

Due to these discontinuities, a single PSF cannot fully describe the image.
Nevertheless, a coadded PSF estimate is required for the lensing analysis. We
must select which PSFs from the individual observations to include in the PSF
coadd.

For the full-image simulation with \mdet\ analysis, we generate a
$300\times300$ coadd image.  We generated a single coadded PSF at the coadd's
center, which is used for subsequent image deconvolution.  While this PSF is
approximate for positions away from the center, it provides sufficient accuracy
for \mdet\ \citep{mdetlsst2023}.  The PSF image from an individual exposure is
included only if the coadd's central pixel is covered by that input image.  If
the central pixel is not covered, the PSF image is excluded from the PSF coadd.

Similarly, for the postage stamp simulations, we include the PSF image from an
input observation in the coadd only if the central pixel of the postage stamp is covered
by that observation.

The same inverse variance weights used for image coaddition were used for PSF
coaddition.

\subsection{Image Processing and Galaxy Measurements}

The images were analyzed using the \mdet\ Python
package\footnote{\url{https://github.com/esheldon/metadetect}}.  We used the
same configuration as in \cite{mdet2020}, with
\texttt{sep}\footnote{\url{https://github.com/sep-developers/sep}} package used
for object detection the
\texttt{ngmix}\footnote{\url{https://github.com/esheldon/ngmix/}} package used
for image shearing and galaxy measurements.  However, unlike in
\cite{mdet2020}, a Gaussian fit was used to approximate the PSF for galaxy
modeling, and a Gaussian fit was used to derive object parameters.  This
Gaussian galaxy model was convolved with the PSF, so that object parameters
such as ellipticity and the trace of the covariance matrix $T$ represent the
``pre-PSF'' galaxy.

Cuts were applied to objects in the realistic simulations to identify well
resolved and relatively high \snr\ objects:  \snr\ $ > 10$ and $\Tratio > 0.5$
(where $T$ is pre-PSF).  A looser cut at $\Tratio > 0$ was also employed.

Note the average \Tratio\ is higher for the simulations with fewer epochs,
because they represent shallower surveys.  Objects that pass the \snr\ cut are
brighter and larger than objects in the deeper simulations.

\section{Results} \label{sec:results}

In the following sections we show results for multiplicative and additive
biases, as well as spatially correlated shape errors.

\subsection{Multiplicative Bias} \label{sec:multbias}

Since our simulations applied a constant weak shear $\vecgam = (0.02, 0.00)$,
the bias parameters \vecm\ and \vecc\ (see Equation \ref{eq:biaslin}) were
extracted by simply averaging the ellipticities in the first and second
ellipticity components, respectively, and applying a mean response correction
as in \cite{mdet2020}.  In no cases did we detect additive biases.  Thus, in
this section, we focus on multiplicative bias only.

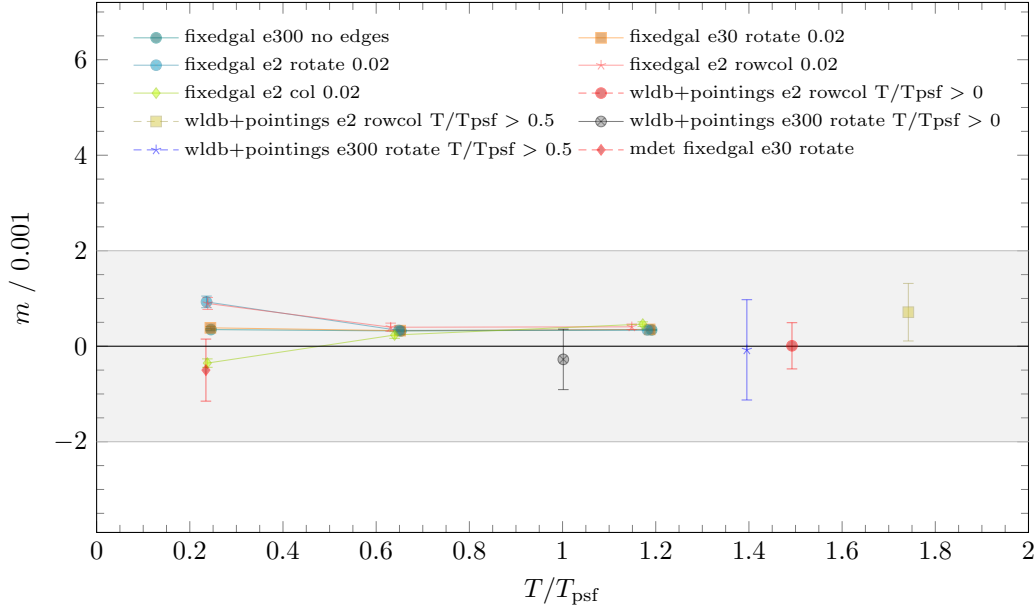
\begin{figure*}[t]
    \centering
    \input{figures/m-vs-Trat-pgf.tex}

    \caption{
        Multiplicative bias $m$ as a function of $T/T^{\mathrm{PSF}}$ for
        various simulation settings.  In the legend, \texttt{e2} indicates two
        epochs, or CCD images, contributing to the coadd, \texttt{e30}
        indicates 30 epochs etc.  \texttt{fixedgal} indicates the fixed
        exponential model, for which all galaxies had a fixed size, and the associated
        value 0.02 (or ``no edges'') indicates the imposed edge hit rate.
        \texttt{wldb+pointings} indicates the realistic simulation with pointings
        intersected to the coadds to create edges, and in which
        the galaxy size is variable.  \texttt{rotate} indicates the input
        images were rotated randomly, such that edges are not purely in the row
        or column direction.  \texttt{rowcol} indicates the images were not
        rotated, and edges may occur in both the row and column directions.
        \texttt{col} indicates the edges were only allowed to be
        present in the column direction.  The point labeled \texttt{mdet
        fixedgal e30 rotate} indicates use of full images and \mdet\ processing
        with detection.
    }

 \label{fig:mvt}

\end{figure*}

In Figure \ref{fig:mvt} we show the multiplicative bias as a function of the
ratio \Tratio\ for various simulation configurations.  We show results for the
cases with 300, 30, and 2 epochs contributing to the coadd on average, and for
both fixed galaxy and realistic galaxy populations.  All simulations use
postage stamps except the one labeled ``mdet fixedgal e30 rotate,'' in which
full coadd images were drawn and full \mdet\ with object detection was
performed.

In all cases the bias is below LSST requirements $m < \lsstreq$
\citep{mdetlsst2023}.  A small bias of order $\sim 4 \times 10^{-4}$ is
expected for the input shear of 0.02 \citep{SheldonMcal2017}.  A negligible,
but significant, departure from the expectation was observed only for very
small galaxies ($\Tratio < 0.6$), which is well below the typical size for
galaxies in lensing samples.  This can be seen by comparing with the realistic
galaxy populations, for which $\Tratio > 1$ even with no explicit cut on
\Tratio.

The small bias above expectations occurred, for the extreme case, both with and
without rotations.  However, with rotations the bias is positive (e.g.
``fixedgal e2 rotate 0.02'') while without rotations the bias is negative (e.g.
``fixedgal e2 col 0.02'').

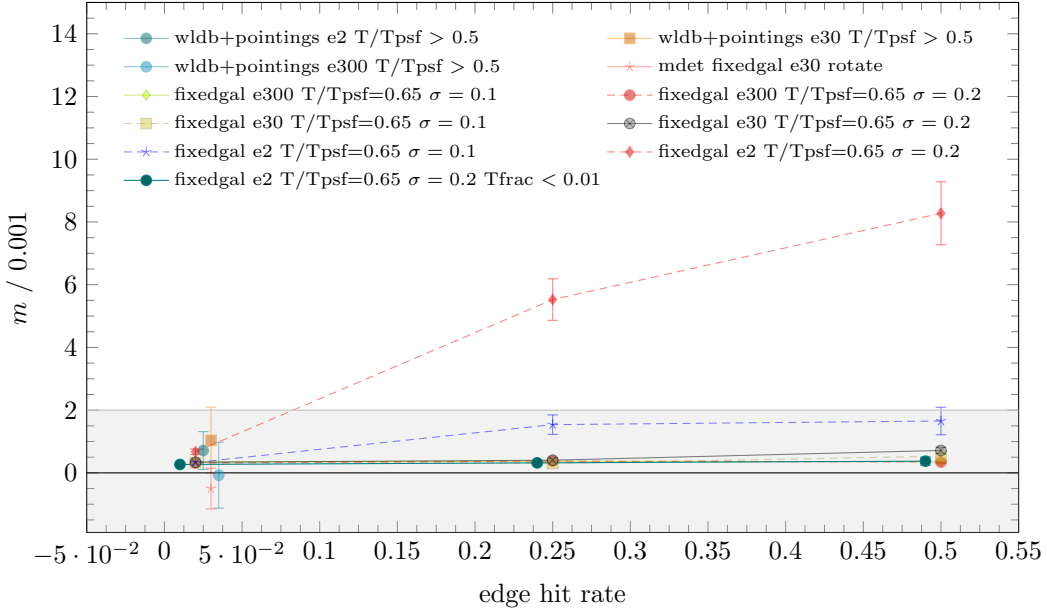
\begin{figure*}[t]
    \centering
    \input{figures/m-vs-edge-hit-rate-pgf.tex}
    \caption{
        Multiplicative bias $m$ as a function of the rate of edge hits in the
        coadd for various simulation settings, including variation of
        the PSF FWHM standard
        deviation $\sigma$.  See figure \ref{fig:mvt} for
        other label definitions.  A bias was found for the case
        of 2 epochs per coadd and $\sigma=0.2$ arcsec.  Cutting
        postage stamps with relatively large PSF size variations due to edges
        \Tfrac=\Tfracdef $< 0.01$ removes the bias.
        The \texttt{wldb+pointings} are artificially offset for clarity, as is
        the data with the \Tfrac\ cut.
    }
    \label{fig:mvr}
\end{figure*}

In Figure \ref{fig:mvr} we show the multiplicative bias as a function of the
rate of edge hits. The edge hit rate is the fraction of images contributing to
the galaxy postage stamp that have an edge in the stamp.  For the simulation
with realistic pointings, the edge hit rate is about 0.03. For the simple
simulations we set the edge hit rate explicitly.   We show results for the
realistic galaxy population as well as one with a fixed $\Tratio = 0.65$ for
all galaxies. As noted above, this \Tratio\ is well below the typical \Tratio\
for real LSST-like galaxy samples.    Two different levels of PSF variations
are shown: $\sigma = 0.1$ and $\sigma = 0.2$.

The bias exceeds LSST requirements only in the case of large PSF
variation and extreme edge hit rate.  We detect a small bias, less than LSST
requirements, for the case of no rotations and high edge hit rate.  The bias is
negligible for all other cases, including the realistic galaxy population.

\subsection{Spatially Correlated Shape Errors} \label{sec:corrbias}

No multiplicative bias (see \S \ref{sec:multbias}) or overall additive bias was
found for typical survey scenarios.  However there may be spatially correlated
additive errors for objects near edges, which would introduce a spurious
into cosmic shear measurements.

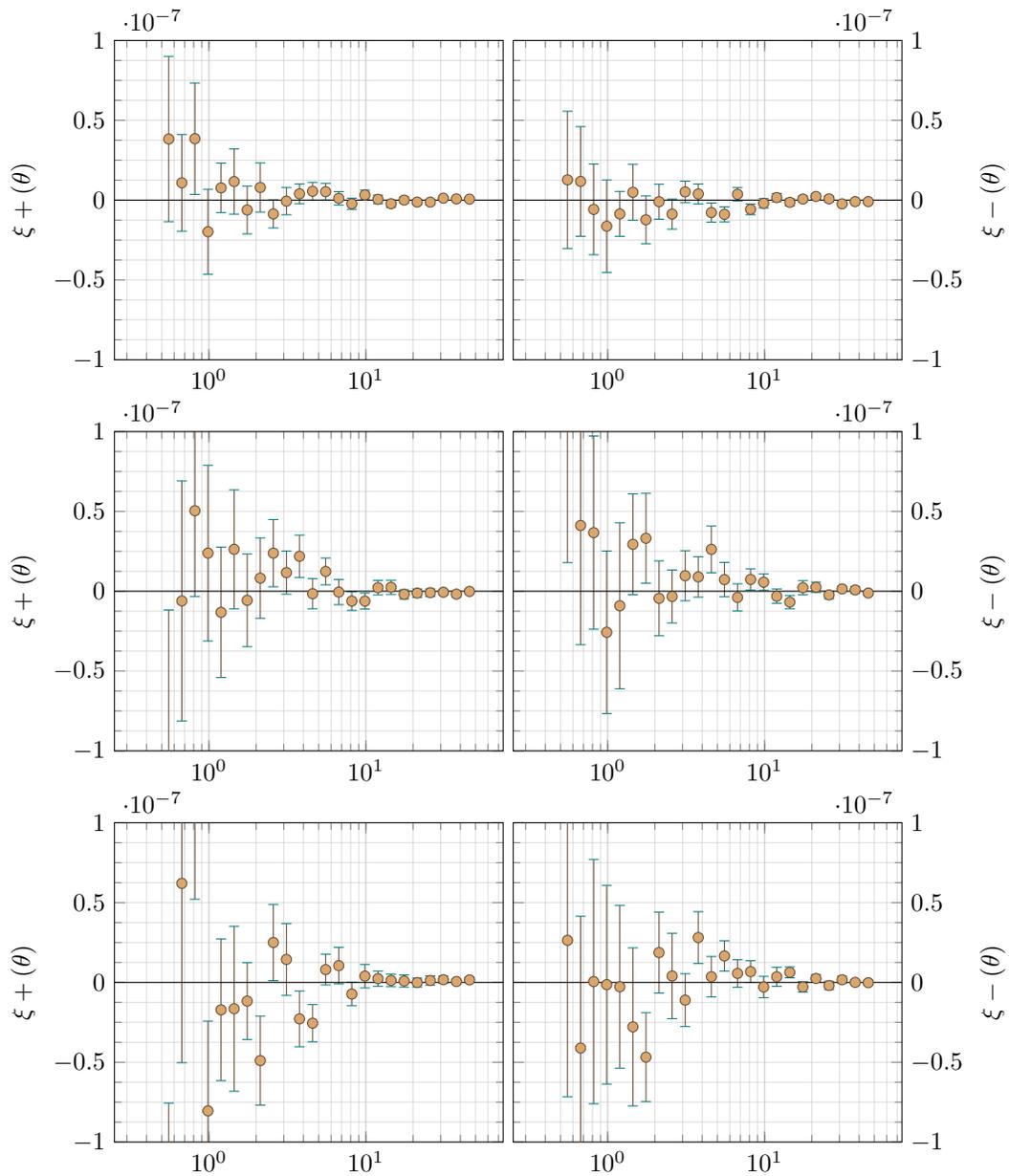
\begin{figure*}[p]
    \centering
    \input{figures/ecorr.tex}
    \caption{
        Correlation functions $\xi+$ and $\xi-$ for zero shear simulations.
        Top row:  A simulation with 300 epochs, on average, contributing to the
        coadd at a given position.  Middle row: A simulation with an average of 30 epochs
        contributing to the coadds.  Bottom row: A simulation with an average
        of 2 epochs contributing to the coadds. No significant correlations were detected,
        and the upper bound on the measurement is well below the expected
        signal in real data.  
    }
    \label{fig:ns}
\end{figure*}

To investigate this potential effect, we repeated the above postage stamp
simulations with zero input shear using the realistic galaxy sample.  In Figure
\ref{fig:ns} we show results for coadds constructed from 300, 30, and 2 epochs,
on average.  These simulations did not include camera rotations; the results
with camera rotations look similar.  In all cases, no significant correlations
were observed in either $\xi+$ or $\xi-$.  Moreover, the measured correlations
are well below the requirements specified for LSST \citep{SRD}.

\subsubsection{Bias in Coadds with Two Epochs}

For a large number of epochs, the absence of net spurious correlations
is expected.  The lack of any detectable effect even for only two epochs,
however, may appear surprising.  We attribute this result to underlying
symmetries in the problem.

First, only a small fraction of objects lie near an edge when two epochs are
combined, which inherently suppresses any potential spurious correlations.

Second, any additive errors arising from PSF inaccuracies must be tied to
specific physical directions within the simulation.  The only relevant
directions are the relative orientation of the PSF between exposures and the
angle of an edge through the coadd.

When camera rotations are applied randomly, the edges do not define a preferred
direction for errors.  Even in the absence of rotations, the required symmetry
is preserved because edges appear in both horizontal and vertical orientations;
any directional errors associated with these axes cancel when the two-point
correlation functions $\xi \pm$ are computed.  Consequently, we focus here on
the relative orientation of the PSFs.

The diffusion kernel is circular and therefore cannot imprint directional
correlations, nor can associated errors.

The atmospheric PSFs are randomly oriented, so the PSF orientations in any
pair of exposures are uncorrelated. If edges-related errors depend on
the relative PSF orientation between exposures, both positive and negative
correlations are expected to occur with equal probability.  These contributions
therefore cancel, producing no net signal in $\xi+$ or $\xi-$.

The optical PSF component has a fixed radial pattern. If pointings were
clustered, and errors were related to the relative orientation of the optical
PSFs at a given location, then a net correlation in the errors could
accumulate.   However, because the pointings are not clustered, we expect no
net contribution to $\xi+/\xi-$ when averaged over many coadds.

The symmetries described above could be broken if pointings exhibit significant
clustering. For areas of sky with few, inhomogeneously distributed pointings,
correlated errors might accumulate. See \S \ref{sec:diagnostic} for a
discussion of diagnostics that could be used to identify and remove problematic
data in such cases.

A persistent wind direction could correlate the PSF ellipticity between
exposures. Associated errors caused by edges might not cancel, although the
effect will be suppressed by the small fraction of objects located near edges
in two image coadds.  Systematic effects in such cases will depend on the
details of the problem, and are thus best studied on an individual basis.

\section{Diagnostic Cuts} \label{sec:diagnostic}

A bias emerges in our simulations for the extreme case of two epochs, high edge
hit rate, and relatively large PSF variations.  In a survey like LSST, this
type of situation occurs only near a survey edge or ``holes'' in the survey
data, such as where bright stars are masked.  A diagnostic to identify objects
in such regions is desirable.

What causes the multiplicative bias?  We do not expect the bias to be due to
discontinuous noise fields, because these can be accommodated by \mcal.  
We will instead focus on diagnosing PSF discontinuities. 

As can be seen by examining Equations \ref{eq:ebias} and \ref{eq:biaslin}, a
multiplicative bias is related to errors in the PSF size. In this work, we fit
a Gaussian model, so the natural size (squared) is the trace of the Gaussian
covariance matrix \Tpsf.  Because of edges, the effective PSF size \Tpsf\ can
vary across the image of a galaxy, but a single PSF must be chosen for lensing
measurements. This mismatch can lead to calibration biases.

To characterize this effect, we can coadd the value of PSF parameters from
images $k$ that contribute to a given pixel $i, j$ and measure
some statistics on the resulting map.  For \Tpsf\ this map is defined as
\begin{equation}
    T_{c}^{\mathrm{PSF}}(i, j) = \frac{\sum_k w_k T^{\mathrm{PSF}}_k(i, j)}{\sum_k w_k},
\end{equation}
where the subscript $c$ indicates the coadded map.  Here $T^{\mathrm{PSF}}_k(i, j)$
is shorthand for the \Tpsf\ from image $k$ as it contributes to pixel $(i, j)$ in the coadd.
In real data, this is evaluated at at the corresponding location in the image
$k$ after it has been warped onto the coadd grid.

In order to diagnose multiplicative biases in particular, we suggest
\begin{equation}
    T_{\mathrm{frac}} = \frac{\sigma(T_c^{\mathrm{PSF}})}{\langle T_c^{\mathrm{PSF}} \rangle}
\end{equation}
the standard deviation over the map \Tpsfmap, divided by the
mean over the map.

The \Tfrac\ statistic should be measured over an area comparable to the size of a
typical object.  We chose to measure the mean and variance within a Gaussian
weighted aperture centered on the object, where the Gaussian weight $G$ is
circular with FWHM = 1 arcsec:
\begin{equation}
    \langle T_c^{\mathrm{PSF}} \rangle = \frac{\sum_{i, j} T_c^{\mathrm{PSF}}(i, j) G(i - x, j - y)}{\sum_{i, j} G(i - x, j - y)},
\end{equation}
where $i, j$ run over all pixel indices and $x$, $y$ are the object center.  A
similar weighted measure of the variance was used.

\begin{figure}[h]
    \centering
    \input{figures/compare-pstats.tex}

    \caption{
        Distribution of \Tfrac=\Tfracdef\ for various simulation
        configurations.  For the realistic LSST-like simulations with 460
        epochs, similar to year 10, 3-band coadds, all values of \Tfrac\ are
        less than 0.002, which is negligible.   For 30 epochs, higher \Tfrac\
        values were found, but only 0.3\% of \Tfrac\ exceeded 0.01.  The more
        extreme, targeted simulations with higher edge hit rate and larger PSF
        variation produced \Tfrac\ distributions with tails to high values.
    }

    \label{fig:tfrac}
\end{figure}
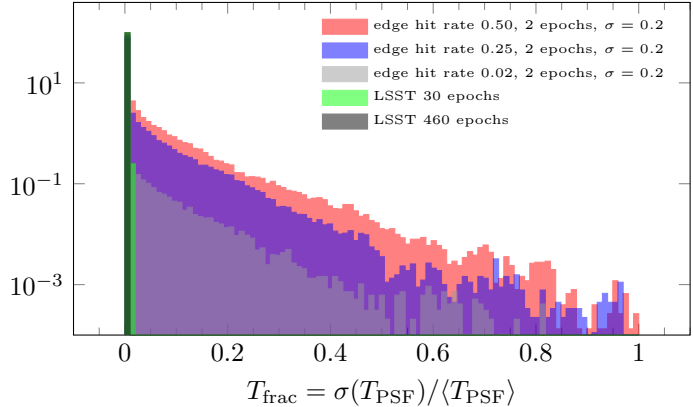

In Figure \ref{fig:tfrac} we show the distribution of \Tfrac\ for various
simulations.  The distribution of \Tfrac\ for the case of 2 epochs and standard
deviation of the PSF FWHM 0.2 arcsec is shown, for three example edge hit
rates.  For these simulations, there is a peak near zero and a broad tail.  We
also show \Tfrac\ values for a simulation with 30 epochs and typical seeing
variations, and an approximate full LSST 10 year simulation with about 460
epochs on average, about what is expected for a combination of $r$, $i$, and $z$
images \citep{SRD}.  For the 30 epoch simulation we again have a peak near
zero, but the tail is small, with a mean value of $3 \times 10^{-4}$ and
maximum value of 0.02.  The \Tfrac\ in the LSST 10 year simulation are all
less than 0.002.

\vspace{1em}

In Figure \ref{fig:mvr}, we show that cutting objects with $\Tfrac > 0.01$
removes the multiplicative bias from the extreme simulations to the expected
level of $\sim 4 \times 10^{-4}$.  Note this cut is
well above the largest value we found for the simulation mimicking LSST year 10
data. For the 30 epoch simulation, the cut removes only 0.3\% of the objects.
It may be that this cut on \Tfrac\ will remove more objects in real surveys,
where there is more unusable data, for example, near bright saturated stars.

Similar statistics can be developed for additive bias (see \S
\ref{sec:corrbias}), for example, by measuring the mean PSF ellipticity in the
Gaussian weighted aperture minus the ellipticity of the PSF used for lensing
measurements.

\section{Discussion} \label{sec:discuss}

Our results indicate that the creation of ``edgeless'' coadds is not necessary
for most imaging survey strategies.  For LSST, the substantive number of images
contributing to each multi-band coadd ($\sim$450 for year 10, $\sim$45 for year
1), and relatively small PSF variations (of order 10\%) ensure that biases
arising from PSF discontinuities are generally negligible.

The simulations with 30 total epochs and no rotational dithers resembles the
Dark Energy Survey year 6, which features three bands ($r, i, z$), $\sim$10
epochs per band and $\sim$arcminute sized coadds.  It is possible that the
approximately \deslost\ of images discarded in the shear analysis
\citep{DESY6Shear2025} could have been retained without introducing additional
bias.  It should be noted, however, that about 10\% (5\%) of objects in the
survey have $\le4$ ($\le3$) input exposures per band in the edge-free coadds
(M. R. Becker, private communication).  A cut on a \Tfrac-like statistic would
remove low exposure count, high PSF variation data that could potentially cause
a bias.

We did not explicitly evaluate a scenario akin to the Hyper Suprime Cam Survey.
In the HSC survey, there are typically 5 to 6 images contributing to the coadd
in $i$-band, and there are no rotational dithers \citep[HSC,][]{HSCDR3}.
However, given the modest biases observed in our 2-epoch simulations, we
anticipate even smaller biases for HSC.  Implementating the \Tfrac\ statistic
would offer further confidence in removing problematic measurements.

A limitation of this study is the lack of ``holes'' in the images, where
certain input data must be excluded prior to coadding.  For instance, near a
bright star, the amount of usable data diminishes towards the star, potentially
leaving no viable data at the center.  Similarly, data from bad amplifiers or
even full CCDs may need to be excluded, which will increase the edge hit rate,
and may be important in regions of sky where the number of visits is low.
Objects in these regions may be subject to substantial PSF variations and thus
high \Tfrac\ values, which could induce significant local shear biases.  To
mitigate such issues, we recommend applying generous masks to exclude objects
near problem regions, ideally in conjunction with a threshold on a statistic
like \Tfrac.

We anticipate that our results are broadly applicable regarding PSF biases.
For noise inhomogeneity, we have demonstrated that \mcal\ and \mdet\ remain
unbiased; however, this may not hold for all shear calibration methods. Our
approach uses a simulated noise field matched to to each image and coadded
using the same weights as the images.  Noise discontinuities are thus naturally
included in the noise images and corrected for in \mdet.  During the
calibration process, the noise field undergoes a 90-degree rotation, followed
by the same deconvolution, shearing, and reconvolution operations applied to
the image. It is then rotated back and added to the image before analysis. This
procedure is intended to eliminate the quadrupole asymmetry in the noise power
spectrum caused by deconvolution and image shearing. Without this correction,
positive and negative sheared images exhibit opposing, non-canceling
asymmetries, leading to an artificially inflated response.

We lack a formal mathematical proof that this noise field method is universally
effective. Other shear calibration algorithms should be tested to confirm that
non-uniform noise effects are adequately addressed.  For methods more sensitive
to non-uniform noise, it may be useful to calculate a \Tfrac-like statistic
over the expected noise in order to remove data with large noise
discontinuities.

In real astronomical data, images are warped prior to coaddition, which
introduces correlations in the noise. For this work we did not warp the images
due to the high computational cost. Thus, we have tested the impact of
discontinuous (but uncorrelated) noise, but not discontinuous correlated noise.

However, in \cite{mdetlsst2023} we demonstrated that \mdet\ accurately
corrects for correlated noise effects when the analysis includes simulated
noise fields that have been processed and coadded in exactly the same
manner as the real data.  Because warping precedes coaddition, we
do not anticipate that the coadded images will contain pathological
noise features that are not already captured by appropriately coadded
simulated noise fields.

Again, a formal proof is not available, but we expect that carefully crafted
noise images, closely matching the actual noise properties of the coadds, can
be used within the \mdet\ framework to provide accurate corrections for noise
inhomogeneities and correlations.


\section{Summary} \label{sec:summary}

We investigated the effects of discontinuous PSF and noise arising from edges
in coadded images on weak shear measurements. Using simple weighted-mean
coaddition, and using \mcal\ and \mdet\ algorithms for shear calibration, we
analyzed both realistic LSST-like simulations and simplified controlled
simulations. Our results demonstrate that for typical galaxy populations,
typical edge-hit rates (a few percent), and coadds formed from tens to hundreds
of input images, no significant multiplicative or additive biases are
introduced, with values well below the requirements for upcoming surveys.
Biases exceeding these requirements occur only in extreme scenarios, such as
coadds with just two input images, an image edge crossing the object in
$\sim$25\% of coadds, PSF size variations greater than 25\%, and relatively
small galaxies.  Even in these cases, we show that such biases can be
effectively mitigated by identifying and removing affected measurements using
the diagnostic statistic \Tfrac=\Tfracdef, which quantifies PSF size variations
across objects.  Furthermore, no spatially correlated shape errors were
detected that could impact cosmic shear analyses.  These findings indicate that
coadded images with edges can be reliably used with \mdet\ in large-scale weak
lensing surveys like LSST, avoiding the need to discard valuable data or
implement noisy PSF homogenization, thereby preserving survey efficiency and
depth.

\section*{Acknowledgments}

Thanks to Jim Bosch for suggesting the author extend a study of missing data in
coadds to include edges.  Thanks to Michael Jarvis for encouraging the
development of a \Tfrac-like statistic. We thank Xiangchong Li and Matt Becker
for helpful comments on the draft of this paper, and Jim Chiang for providing
the operation simulation from DC2.  We thank the excellent computing staff of
the RHIC Atlas Computing Facility at Brookhaven National Laboratory and the
SLAC Shared Science Data Facility for their support.  Finally, many thanks to
the referees whose careful reading made this paper substantially better.  ES is
supported by DOE grant DE-AC02-98CH10886.

\let\bibsection\relax
\section*{\centering References}


\footnotesize
\bibliographystyle{aasjournal}
\bibliography{references}

\end{document}

%% file: commands.tex
\newcommand{\Tpsf}{$T^{\mathrm{PSF}}$}

\newcommand{\Tpsfmap}{$T_c^{\mathrm{PSF}}(i, j)$}
\newcommand{\Tfracdef}{$\sigma(T_c^{\mathrm{PSF}})/\langle T_c^{\mathrm{PSF}} \rangle$}
\newcommand{\Tfrac}{\mbox{$T_{\mathrm{frac}}$}}

\newcommand{\descwl}{\texttt{WeakLensingDeblending}}

\newcommand{\vecc}{\mbox{\boldmath $c$}}
\newcommand{\vecm}{\mbox{\boldmath $m$}}

\newcommand{\galsim}{\texttt{GALSIM}}

\newcommand{\snr}{$S/N$}
\newcommand{\Tratio}{\mbox{$T/T_{\mathrm{PSF}}$}}
\newcommand{\lsstreq}{0.002}

\newcommand{\deslost}{25\%}

\newcommand{\mcal}{\textsc{metacalibration}}
\newcommand{\mdet}{\textsc{metadetection}}

\newcommand{\gest}{\mbox{\boldmath $\hat \gamma$}}
\newcommand{\vecgam}{\mbox{\boldmath $\gamma$}}

\newcommand{\imsim}{\texttt{imSim}}
\newcommand{\descwlshearsims}{\texttt{descwl-shear-sims}}

%% file: colors.tex
\definecolor{bisque}{HTML}{ffe4c4}
\definecolor{bisque}{HTML}{ffe8c4}
\definecolor{tan}{HTML}{d2b48c}
\definecolor{sand}{HTML}{e1bf92}
\definecolor{sand1}{HTML}{f6d7b0}
\definecolor{sand2}{HTML}{f2d2a9}
\definecolor{sand3}{HTML}{eccca2}
\definecolor{sand4}{HTML}{e7c496}
\definecolor{sand5}{HTML}{e1bf92}

\definecolor{desert1}{HTML}{d9a76c}
\definecolor{desert2}{HTML}{a48963}
\definecolor{desert3}{HTML}{674c37}
\definecolor{desert4}{HTML}{512f26}
\definecolor{desert5}{HTML}{211813}

\definecolor{bsand1}{HTML}{b18e5d}
\definecolor{bsand2}{HTML}{97744f}
\definecolor{bsand3}{HTML}{590d0d}
\definecolor{bsand4}{HTML}{3b0808}
\definecolor{bsand5}{HTML}{1d0808}

%% file: figures/psf-fwhm.tex
\begin{tikzpicture}
\begin{axis} [minor tick num=3,cycle list name=exotic,xlabel={PSF FWHM [arcsec]},width=1.0\columnwidth,height=0.614\columnwidth,legend style={draw=none,fill=none,legend columns=1,legend cell align=left},ymin=0]
\addplot+ [no markers,opacity=0.5,brown,fill,draw,const plot mark right] table  {
x y
0.5610555410385132 0.0
0.5795113444328308 1.0
0.5979670882225037 6.0
0.6164228916168213 15.0
0.6348786950111389 47.0
0.6533344984054565 96.0
0.6717902421951294 215.0
0.690246045589447 319.0
0.7087018489837646 427.0
0.7271575927734375 592.0
0.7456133961677551 760.0
0.7640691995620728 883.0
0.7825249433517456 938.0
0.8009807467460632 951.0
0.8194365501403809 938.0
0.8378922939300537 858.0
0.8563480973243713 780.0
0.874803900718689 676.0
0.8932596445083618 550.0
0.9117154479026794 453.0
0.9301712512969971 366.0
0.9486269950866699 287.0
0.9670827984809875 244.0
0.9855386018753052 158.0
1.003994345664978 131.0
1.0224502086639404 96.0
1.0409059524536133 58.0
1.0593616962432861 54.0
1.077817440032959 20.0
1.0962733030319214 26.0
1.1147291660308838 12.0
1.1331849098205566 11.0
1.1516406536102295 14.0
1.1700963973999023 4.0
1.1885521411895752 9.0
1.2070080041885376 1.0
1.2254638671875 3.0
1.2439196109771729 0.0
1.2623753547668457 0.0
1.2808310985565186 0.0
1.299286961555481 1.0
1.299286961555481 0.0
};
\node[anchor=mid west] at (axis description cs:0.55,0.9) {mean: 0.81 std: 0.09};
\end{axis}
\end{tikzpicture}

%% file: figures/m-vs-Trat-pgf.tex
\begin{tikzpicture}
\begin{axis} [minor tick num=3,cycle list name=exotic,xlabel={$T/T_{\mathrm{psf}}$},ylabel={$m~/~0.001$},xmin=0,xmax=2,ymin=-3.9,ymax=7.2,width=0.9\textwidth,height=0.55\textwidth,legend style={draw=none,fill=none,legend columns=2,legend cell align=left,legend pos=north west,font=\scriptsize}]
\addplot+ [opacity=0.5,error bars/.cd,error bar style={solid},y dir=both,y explicit] table [y error=yerr] {
x y yerr
0.24536751276345958 0.3482611886336251 0.021063825721248625
0.6530845607472557 0.3197032037620673 0.014187392352027862
1.191262329122743 0.3411837341185109 0.009999414693356476
};
\addlegendentry{{fixedgal e300 no edges}}
\addplot+ [opacity=0.5,error bars/.cd,error bar style={solid},y dir=both,y explicit] table [y error=yerr] {
x y yerr
0.24416885011379535 0.38524649124394017 0.0284110382162803
0.6514768319200663 0.3260992390412554 0.014504056994736234
1.1868633504463995 0.3520540584671661 0.009199812202384076
};
\addlegendentry{{fixedgal e30 rotate 0.02}}
\addplot+ [opacity=0.5,error bars/.cd,error bar style={solid},y dir=both,y explicit] table [y error=yerr] {
x y yerr
0.23584284002905284 0.9285534750944624 0.12197511881118604
0.648626998554432 0.33112211083796694 0.011746869636248108
1.1821118798165684 0.34133927137669495 0.008923867527934276
};
\addlegendentry{{fixedgal e2 rotate 0.02}}
\addplot+ [opacity=0.5,error bars/.cd,error bar style={solid},y dir=both,y explicit] table [y error=yerr] {
x y yerr
0.23839324967696907 0.8956273738363052 0.1253351866099405
0.6311221159384844 0.39780684723123017 0.08598315958025367
1.1486244512960717 0.4062165045353172 0.057798497615391596
};
\addlegendentry{{fixedgal e2 rowcol 0.02}}
\addplot+ [opacity=0.5,error bars/.cd,error bar style={solid},y dir=both,y explicit] table [y error=yerr] {
x y yerr
0.2376103700258669 -0.35342750387412636 0.08852793628297534
0.6395443902928986 0.23486912662962212 0.07015388922088733
1.1722174836712569 0.46287045593040865 0.04819673114031701
};
\addlegendentry{{fixedgal e2 col 0.02}}
\addplot+ [opacity=0.5,error bars/.cd,error bar style={solid},y dir=both,y explicit] table [y error=yerr] {
x y yerr
1.4923896189720451 0.008956940240611999 0.48390851164431653
};
\addlegendentry{{wldb+pointings e2 rowcol T/Tpsf $>$ 0}}
\addplot+ [opacity=0.5,error bars/.cd,error bar style={solid},y dir=both,y explicit] table [y error=yerr] {
x y yerr
1.7421477638444787 0.7124103158304695 0.602543685326254
};
\addlegendentry{{wldb+pointings e2 rowcol T/Tpsf $>$ 0.5}}
\addplot+ [opacity=0.5,error bars/.cd,error bar style={solid},y dir=both,y explicit] table [y error=yerr] {
x y yerr
1.0016440160152575 -0.2755485770946642 0.6328491938080719
};
\addlegendentry{{wldb+pointings e300 rotate T/Tpsf $>$ 0}}
\addplot+ [opacity=0.5,error bars/.cd,error bar style={solid},y dir=both,y explicit] table [y error=yerr] {
x y yerr
1.3955605553836563 -0.07574368248441932 1.050280648685681
};
\addlegendentry{{wldb+pointings e300 rotate T/Tpsf $>$ 0.5}}
\addplot+ [opacity=0.5,error bars/.cd,error bar style={solid},y dir=both,y explicit] table [y error=yerr] {
x y yerr
0.23458818992406566 -0.5002875645757632 0.6499939555841865
};
\addlegendentry{{mdet fixedgal e30 rotate}}

\addplot[fill=lightgray,opacity=0.2,draw=none,area legend] coordinates {(0,-2) (2,-2) (2,2)(0,2)};
\addlegendentry{LSST Requirement}

\addplot+ [no markers,black,solid] table  {
x y
0 0
2 0
};
\addplot+ [no markers,lightgray,solid,name path=upper] table  {
x y
0 2
2 2
};
\addplot+ [no markers,lightgray,solid,name path=lower] table  {
x y
0 -2
2 -2
};


\end{axis}
\end{tikzpicture}

%% file: figures/m-vs-edge-hit-rate-pgf.tex
\begin{tikzpicture}
\begin{axis} [minor tick num=3,cycle list name=exotic,xlabel={edge hit rate},ylabel={$m~/~0.001$},xmin=-0.05,xmax=0.55,ymin=-1.9,ymax=15.0,width=0.9\textwidth,height=0.55\textwidth,legend style={draw=none,fill=none,legend columns=2,legend cell align=left,legend pos=north west,font=\scriptsize}]
\addplot+ [opacity=0.5,error bars/.cd,error bar style={solid},y dir=both,y explicit] table [y error=yerr] {
x y yerr
0.025 0.7124103158304695 0.602543685326254
};
\addlegendentry{{wldb+pointings e2 T/Tpsf $>$ 0.5}}
\addplot+ [opacity=0.5,error bars/.cd,error bar style={solid},y dir=both,y explicit] table [y error=yerr] {
x y yerr
0.03 1.0371762767802384 1.0568679293487846
};
\addlegendentry{{wldb+pointings e30 T/Tpsf $>$ 0.5}}
\addplot+ [opacity=0.5,error bars/.cd,error bar style={solid},y dir=both,y explicit] table [y error=yerr] {
x y yerr
0.035 -0.07574368248441932 1.050280648685681
};
\addlegendentry{{wldb+pointings e300 T/Tpsf $>$ 0.5}}
\addplot+ [opacity=0.5,error bars/.cd,error bar style={solid},y dir=both,y explicit] table [y error=yerr] {
x y yerr
0.03 -0.5002875645757632 0.6499939555841865
};
\addlegendentry{{mdet fixedgal e30 rotate}}
\addplot+ [opacity=0.5,error bars/.cd,error bar style={solid},y dir=both,y explicit] table [y error=yerr] {
x y yerr
0.02 0.35427020661304987 0.013732942226163608
0.25 0.3710063789750251 0.02239716366251393
0.5 0.354757412338369 0.03569464253983878
};
\addlegendentry{{fixedgal e300 T/Tpsf=0.65 $\sigma=0.1$}}
\addplot+ [opacity=0.5,error bars/.cd,error bar style={solid},y dir=both,y explicit] table [y error=yerr] {
x y yerr
0.02 0.3114077181871622 0.02017519882003238
0.25 0.37558176518093234 0.024461781488087888
0.5 0.34639028040261977 0.0383070960467524
};
\addlegendentry{{fixedgal e300 T/Tpsf=0.65 $\sigma=0.2$}}
\addplot+ [opacity=0.5,error bars/.cd,error bar style={solid},y dir=both,y explicit] table [y error=yerr] {
x y yerr
0.02 0.3260992390412554 0.014504056994736234
0.25 0.2998779322347378 0.05560440676958647
0.5 0.5306355751819591 0.09588573481237615
};
\addlegendentry{{fixedgal e30 T/Tpsf=0.65 $\sigma=0.1$}}
\addplot+ [opacity=0.5,error bars/.cd,error bar style={solid},y dir=both,y explicit] table [y error=yerr] {
x y yerr
0.02 0.3383186528644888 0.017168632175964125
0.25 0.4047175787456059 0.06124276360564297
0.5 0.7142020481056832 0.10457199790681258
};
\addlegendentry{{fixedgal e30 T/Tpsf=0.65 $\sigma=0.2$}}
\addplot+ [opacity=0.5,error bars/.cd,error bar style={solid},y dir=both,y explicit] table [y error=yerr] {
x y yerr
0.02 0.33112211083796694 0.011746869636248108
0.25 1.5376156163668941 0.30997337035975514
0.5 1.6534099200880004 0.43790570716938343
};
\addlegendentry{{fixedgal e2 T/Tpsf=0.65 $\sigma=0.1$}}
\addplot+ [opacity=0.5,error bars/.cd,error bar style={solid},y dir=both,y explicit] table [y error=yerr] {
x y yerr
0.02 0.6743118092114031 0.08652261820345855
0.25 5.5253205959238105 0.6618073196228524
0.5 8.277316061114703 1.0037517124351807
};
\addlegendentry{{fixedgal e2 T/Tpsf=0.65 $\sigma=0.2$}}
\addplot+ [opacity=1.0,error bars/.cd,error bar style={solid},y dir=both,y explicit] table [y error=yerr] {
x y yerr
0.01 0.2676772280933992 0.01661601056293049
0.24 0.3189371635312632 0.07317398507122803
0.49 0.3728924007719847 0.12601800042290676
};
\addlegendentry{{fixedgal e2 T/Tpsf=0.65 $\sigma=0.2$ Tfrac $< 0.01$}}

\addplot[fill=lightgray,opacity=0.2,draw=none,area legend] coordinates {(-0.05,-2) (0.55,-2) (0.55,2)(-0.05,2)};
\addlegendentry{LSST Requirement}

\addplot+ [no markers,black,solid] table  {
x y
-0.05 0
0.55 0
};
\addplot+ [no markers,lightgray,solid,name path=upper] table  {
x y
-0.05 2
0.55 2
};
\addplot+ [no markers,lightgray,solid,name path=lower] table  {
x y
-0.05 -2
0.55 -2
};
\end{axis}
\end{tikzpicture}

%% file: figures/ecorr.tex
\begin{tikzpicture}
\begin{groupplot} [minor tick num=3,xmode=log,cycle list name=exotic,xmin=0.25,xmax=75.0,ymin=-1e-07,ymax=1e-07,grid=both,grid style={color=lightgray,opacity=0.5},legend style={draw=none,fill=none,legend columns=1,legend cell align=left},group style={group size=2 by 3,horizontal sep=0.01\columnwidth},width=0.9\columnwidth]
\nextgroupplot [minor tick num=3,cycle list name=exotic,legend style={draw=none,fill=none,legend columns=1,legend cell align=left},ylabel={$\xi+(\theta)$}]
\addplot+[
    only marks,
    mark=*,
    mark options={draw=desert3, fill=desert1}, 
    error bars/.cd,
    error bar style={solid,draw=desert3},
    y dir=both,
    y explicit
] table [y error=yerr] {
x y yerr
0.5545578894143292 3.82857414666151e-08 5.177454267536657e-08
0.6718622646805608 1.0832072742459097e-08 3.0297676499986225e-08
0.8139532039832217 3.84786257420206e-08 3.490925216702129e-08
0.9861589923916481 -1.9817408133591544e-08 2.661125662178638e-08
1.1947209154686436 7.65786285409242e-09 1.549755450091747e-08
1.4474257847067225 1.1637189511864787e-08 2.0393410759165913e-08
1.7535922699008273 -6.187490427577316e-09 1.4967800950365575e-08
2.1244917488325905 7.922982273687464e-09 1.5412563085417784e-08
2.5738416623686122 -8.641665700579612e-09 8.842620098590095e-09
3.11817297141765 -6.391890115450227e-10 8.5467232922321e-09
3.777664081804386 3.962254316598191e-09 6.176941129320558e-09
4.576758594575576 5.531850005202367e-09 5.523708343181035e-09
5.544902853819703 5.272049030272447e-09 5.210109684105078e-09
6.717864258579822 1.103970424615782e-09 4.228012050156238e-09
8.13895059076526 -2.271518112026947e-09 3.5010246931885765e-09
9.860418230810124 3.2466506073098293e-09 3.0119728967192602e-09
11.946107418802274 4.82505410764002e-10 2.6895059450501505e-09
14.472697919162298 -2.2583845705447837e-09 2.2784216072703375e-09
17.534057272123178 1.75867233826937e-11 1.9344074885830466e-09
21.241809689067033 -1.226756018241591e-09 1.6630998210813987e-09
25.73412393775297 -1.2435753144698203e-09 1.534161766543209e-09
31.17518927142336 1.2039018346553157e-09 1.1436050254874617e-09
37.7689800886934 7.608370992930674e-10 9.413759765398845e-10
45.752084025608646 5.964566481590058e-10 1.1621020287196564e-09
};
\addplot+[no markers,black,samples=100,trig format=rad,domain=0.25:75.0] { 0 };
\nextgroupplot [minor tick num=3,cycle list name=exotic,legend style={draw=none,fill=none,legend columns=1,legend cell align=left},ylabel={$\xi-(\theta)$},ylabel near ticks,yticklabel pos=right]
\addplot+[
    only marks,
    mark=*,
    mark options={draw=desert3, fill=desert1}, 
    error bars/.cd,
    error bar style={solid,draw=desert3},
    y dir=both,
    y explicit
] table [y error=yerr] {
x y yerr
0.5545578894143292 1.2685849502943944e-08 4.2978179722270245e-08
0.6718622646805608 1.1698699897038071e-08 3.434774083218568e-08
0.8139532039832217 -5.808935665257014e-09 2.8423008043715372e-08
0.9861589923916481 -1.6357840961943632e-08 2.8999934225318996e-08
1.1947209154686436 -8.62358803315928e-09 1.407152343191249e-08
1.4474257847067225 4.899202703581638e-09 1.7502568306883924e-08
1.7535922699008273 -1.2312723850237668e-08 1.497716909439625e-08
2.1244917488325905 -9.858588208743035e-10 1.0956905609810559e-08
2.5738416623686122 -8.744510092368173e-09 9.462316051790167e-09
3.11817297141765 5.159533574021092e-09 6.702287914220466e-09
3.777664081804386 3.874373575935176e-09 6.231531329528726e-09
4.576758594575576 -7.779882286758866e-09 5.974719049597889e-09
5.544902853819703 -8.888155166221944e-09 4.708576379837558e-09
6.717864258579822 3.684884378336427e-09 4.211579139392625e-09
8.13895059076526 -5.7985599624522254e-09 3.3687570556708857e-09
9.860418230810124 -2.015319583187023e-09 2.9694548998991905e-09
11.946107418802274 1.5770391544676812e-09 2.6017944895565058e-09
14.472697919162298 -1.3912378597772215e-09 2.2495055205660297e-09
17.534057272123178 6.794399687496801e-10 1.8474729807582263e-09
21.241809689067033 2.292009456733936e-09 1.7090616729159058e-09
25.73412393775297 7.743643121679957e-10 1.7808839653609979e-09
31.17518927142336 -2.341725089927966e-09 9.471078367084692e-10
37.7689800886934 -9.259096333261381e-10 9.144371872764874e-10
45.752084025608646 -9.199355166032259e-10 1.0755854816221735e-09
};
\addplot+[no markers,black,samples=100,trig format=rad,domain=0.25:75.0] { 0 };
\nextgroupplot [minor tick num=3,cycle list name=exotic,legend style={draw=none,fill=none,legend columns=1,legend cell align=left},ylabel={$\xi+(\theta)$}]
\addplot+[
    only marks,
    mark=*,
    mark options={draw=desert3, fill=desert1}, 
    error bars/.cd,
    error bar style={solid,draw=desert3},
    y dir=both,
    y explicit
] table [y error=yerr] {
x y yerr
0.5545704093963324 -1.0307885752116557e-07 9.129574895783563e-08
0.6718774127123452 -6.076824211155724e-09 7.515688753545499e-08
0.8139468781913692 5.042432932637446e-08 5.3655626431044526e-08
0.9861455978296673 2.388610101090514e-08 5.505437965109094e-08
1.194719502071013 -1.319118731721697e-08 4.085272732605426e-08
1.447411416027643 2.628316743061693e-08 3.726919146007894e-08
1.7535347719966818 -5.676302025790653e-09 2.9016223382083375e-08
2.1244560820223697 8.227374304267613e-09 2.5148081184868032e-08
2.5738120213783247 2.386513163842989e-08 2.103578406628602e-08
3.11815460217457 1.1680435646481204e-08 1.350234223623123e-08
3.7776674720296817 2.1897523795730304e-08 1.3232562694609255e-08
4.5767206655618535 -1.5247233452946279e-09 9.444590620788092e-09
5.544856447313714 1.2415268981263928e-08 8.384493452078313e-09
6.717888649736104 -5.069227605597526e-10 7.898937375996555e-09
8.138968908126266 -6.130537141219907e-09 5.87020961314365e-09
9.860505539850799 -6.172904361615317e-09 5.0033357538201795e-09
11.946327891832436 2.2440345289498465e-09 4.569415237272456e-09
14.473332944862614 2.4948185622385952e-09 4.4383736049575785e-09
17.53500947336923 -1.927896726008208e-09 2.936021343805089e-09
21.243348212071567 -1.1537745996914904e-09 2.6593295587171763e-09
25.736203718373982 -8.958369228016524e-10 2.409562304407471e-09
31.17923121682671 -6.169154810080063e-10 1.5295860832487585e-09
37.77398176583618 -1.8164988910829907e-09 1.866990992731756e-09
45.76051958736929 -1.214333432352259e-10 1.4132461249512123e-09
};
\addplot+[no markers,black,samples=100,trig format=rad,domain=0.25:75.0] { 0 };
\nextgroupplot [minor tick num=3,cycle list name=exotic,legend style={draw=none,fill=none,legend columns=1,legend cell align=left},ylabel={$\xi-(\theta)$},ylabel near ticks,yticklabel pos=right]
\addplot+[
    only marks,
    mark=*,
    mark options={draw=desert3, fill=desert1}, 
    error bars/.cd,
    error bar style={solid,draw=desert3},
    y dir=both,
    y explicit
] table [y error=yerr] {
x y yerr
0.5545704093963324 1.1533451455173632e-07 9.742097957116388e-08
0.6718774127123452 4.120477463722844e-08 7.468682681661318e-08
0.8139468781913692 3.671812823355932e-08 6.051763650509922e-08
0.9861455978296673 -2.5705685836774813e-08 5.0906031179430476e-08
1.194719502071013 -9.088420950120523e-09 5.1983988123855e-08
1.447411416027643 2.9410148808095138e-08 3.1558815136645376e-08
1.7535347719966818 3.320493951758841e-08 2.8147896735085987e-08
2.1244560820223697 -4.450366197276131e-09 2.3463125538930953e-08
2.5738120213783247 -3.270783661297099e-09 1.657821408478163e-08
3.11815460217457 9.715613829847805e-09 1.5679445759842738e-08
3.7776674720296817 8.987119291061144e-09 1.264820260548406e-08
4.5767206655618535 2.6254824216964524e-08 1.4654701237004736e-08
5.544856447313714 7.3121066395423125e-09 1.0772067904599292e-08
6.717888649736104 -3.86570602259625e-09 8.539533641325686e-09
8.138968908126266 7.393705612659955e-09 6.674338403546499e-09
9.860505539850799 5.698568154196262e-09 5.124998175421474e-09
11.946327891832436 -3.109909230665279e-09 4.474678864470642e-09
14.473332944862614 -6.7947824110665855e-09 4.172807363310095e-09
17.53500947336923 2.2142311805178e-09 4.386372191530566e-09
21.243348212071567 2.533163327961776e-09 3.2287499004024035e-09
25.736203718373982 -2.2779278054971424e-09 2.484675398297282e-09
31.17923121682671 1.4740799841871722e-09 1.7452387518858416e-09
37.77398176583618 8.430681382581695e-10 2.085551552334203e-09
45.76051958736929 -1.1102798688010518e-09 1.4158713878999334e-09
};
\addplot+[no markers,black,samples=100,trig format=rad,domain=0.25:75.0] { 0 };
\nextgroupplot [minor tick num=3,cycle list name=exotic,legend style={draw=none,fill=none,legend columns=1,legend cell align=left},ylabel={$\xi+(\theta)$}]
\addplot+[
    only marks,
    mark=*,
    mark options={draw=desert3, fill=desert1}, 
    error bars/.cd,
    error bar style={solid,draw=desert3},
    y dir=both,
    y explicit
] table [y error=yerr] {
x y yerr
0.5545376290664713 -1.7657367355023648e-07 1.0101394113134144e-07
0.671822383049528 6.203988903416274e-08 1.1237536379370994e-07
0.813949104636567 1.26226053043066e-07 7.414795090294533e-08
0.9860866289058144 -8.048557533208893e-08 5.624373415591443e-08
1.1947257916554674 -1.7215384542434853e-08 4.438235173848534e-08
1.4474199736014781 -1.650088804967298e-08 5.169881860224021e-08
1.7535304006385561 -1.1724162454380286e-08 2.4040986361640286e-08
2.1244536665465352 -4.89914058775472e-08 2.7855364083279896e-08
2.5737714001139955 2.49769072026509e-08 2.3900945145960915e-08
3.1181191350057627 1.4372946331179435e-08 2.252488978979409e-08
3.777621192303385 -2.284020325781816e-08 1.7473058743191445e-08
4.576754285726625 -2.5551562663792204e-08 1.1658413570858785e-08
5.544995229715258 8.009359840720573e-09 9.599936520615236e-09
6.71801242545696 1.0476401909107384e-08 1.1449120084245939e-08
8.139022250302421 -7.29649437630315e-09 7.325800804278191e-09
9.860707867416336 3.929539218234732e-09 7.365394090640504e-09
11.946458517579186 2.3492363265539143e-09 4.773152500724854e-09
14.473904195971643 1.1963269765534564e-09 3.914258156276864e-09
17.534615749489163 8.357498628327471e-10 3.786320434379696e-09
21.24350710171422 -1.0987188887724663e-10 2.7819864181049698e-09
25.737071173017476 1.1657625700843652e-09 2.8860587510705854e-09
31.180169031667198 1.6714017157427713e-09 2.2207581139891006e-09
37.77484879187491 5.297368502907737e-10 1.6739861156866537e-09
45.7646812665937 1.5523569933837354e-09 1.7626660145069085e-09
};
\addplot+[no markers,black,samples=100,trig format=rad,domain=0.25:75.0] { 0 };
\nextgroupplot [minor tick num=3,cycle list name=exotic,legend style={draw=none,fill=none,legend columns=1,legend cell align=left},ylabel={$\xi-(\theta)$},ylabel near ticks,yticklabel pos=right]
\addplot+[
    only marks,
    mark=*,
    mark options={draw=desert3, fill=desert1}, 
    error bars/.cd,
    error bar style={solid,draw=desert3},
    y dir=both,
    y explicit
] table [y error=yerr] {
x y yerr
0.5545376290664713 2.6352117415921024e-08 9.807518287896999e-08
0.671822383049528 -4.113820134378354e-08 8.255373094623051e-08
0.813949104636567 4.907920852463015e-10 7.652982594276523e-08
0.9860866289058144 -1.4181859616898814e-09 6.223441197795785e-08
1.1947257916554674 -2.753694059013878e-09 5.1054357710217254e-08
1.4474199736014781 -2.7858798798751315e-08 4.949246705642483e-08
1.7535304006385561 -4.6830687650099415e-08 2.788290871878025e-08
2.1244536665465352 1.867362859257757e-08 2.5405647018113405e-08
2.5737714001139955 3.992949286050031e-09 2.671686790782545e-08
3.1181191350057627 -1.1099932248927654e-08 1.6546894189202347e-08
3.777621192303385 2.8081666820980638e-08 1.6215416416803367e-08
4.576754285726625 3.5499799812857366e-09 1.267472219730012e-08
5.544995229715258 1.6552409939610376e-08 9.468280916992769e-09
6.71801242545696 5.610349408823122e-09 8.637681132522798e-09
8.139022250302421 6.7370238854634195e-09 6.855371214420391e-09
9.860707867416336 -2.9078745808883166e-09 6.668802003875076e-09
11.946458517579186 3.5044666402161302e-09 5.981055430733852e-09
14.473904195971643 6.273242798670308e-09 3.419262714586331e-09
17.534615749489163 -2.8552294122533963e-09 3.2439086964254e-09
21.24350710171422 2.437349982350257e-09 2.4076879045891605e-09
25.737071173017476 -1.937790131909666e-09 2.615229341299252e-09
31.180169031667198 1.6337096533816527e-09 2.1546784832218806e-09
37.77484879187491 6.645757588226714e-12 1.6085159873733854e-09
45.7646812665937 -1.972697961133465e-10 1.4670832415569324e-09
};
\addplot+[no markers,black,samples=100,trig format=rad,domain=0.25:75.0] { 0 };
\end{groupplot}
\end{tikzpicture}

%% file: figures/compare-pstats.tex
\begin{tikzpicture}
\begin{axis} [minor tick num=3,ymode=log,cycle list name=exotic,xlabel={$T_{\mathrm{frac}} = \sigma(T_{\mathrm{PSF}}) / \langle T_{\mathrm{PSF}} \rangle$},ymin=0.0001,width=1.0\columnwidth,height=0.614\columnwidth,legend style={draw=none,fill=none,legend columns=1,legend cell align=left,area legend,font=\tiny}]
\addplot+ [no markers,opacity=0.5,red,fill,draw,const plot mark right] table  {
x y
0.0 0.0001
0.010101010101010102 75.37375378569199
0.020202020202020204 4.383093946777289
0.030303030303030304 2.8085059358846767
0.04040404040404041 2.0568834964590965
0.05050505050505051 1.7184677474446157
0.06060606060606061 1.4342886917478952
0.07070707070707072 1.3028741120953264
0.08080808080808081 1.1246488900438147
0.09090909090909091 0.9267976762782952
0.10101010101010102 0.8557197603410832
0.11111111111111112 0.7274877739394373
0.12121212121212122 0.6540229130640366
0.13131313131313133 0.6198100356166469
0.14141414141414144 0.5033801658538279
0.15151515151515152 0.48826284791195984
0.16161616161616163 0.41440016288019293
0.17171717171717174 0.34822874487148375
0.18181818181818182 0.29160510661553496
0.19191919191919193 0.28099646244580123
0.20202020202020204 0.2489053138323591
0.21212121212121213 0.23604233277655853
0.22222222222222224 0.1662904973605638
0.23232323232323235 0.16509702489146885
0.24242424242424243 0.13632107758106865
0.25252525252525254 0.13804498225864986
0.26262626262626265 0.13539282121621662
0.27272727272727276 0.1282319864016469
0.2828282828282829 0.10210820013367954
0.29292929292929293 0.11192119599068312
0.30303030303030304 0.09136694791182494
0.31313131313131315 0.08433872114937686
0.32323232323232326 0.07412790113600891
0.33333333333333337 0.07094530788508903
0.3434343434343435 0.0672322824256825
0.3535353535353536 0.06046927176747775
0.36363636363636365 0.04972801954562342
0.37373737373737376 0.05184974837956974
0.38383838383838387 0.052910612796543036
0.393939393939394 0.04826933097228487
0.4040404040404041 0.036467214333456975
0.4141414141414142 0.032223756665563805
0.42424242424242425 0.03156071640495566
0.43434343434343436 0.038854159271646885
0.4444444444444445 0.03872155121952523
0.4545454545454546 0.029041163414643918
0.4646464646464647 0.02267597691280416
0.4747474747474748 0.02439988159038576
0.48484848484848486 0.018697735349154407
0.494949494949495 0.01684122261945104
0.5050505050505051 0.012862981055801189
0.5151515151515152 0.012199940795192814
0.5252525252525253 0.012862981055801258
0.5353535353535354 0.009945603909124684
0.5454545454545455 0.008486915335786304
0.5555555555555556 0.006365186501839798
0.5656565656565657 0.00795648312729966
0.5757575757575758 0.007956483127299747
0.5858585858585859 0.00861952338790806
0.595959595959596 0.005967362345474745
0.6060606060606061 0.0074260509188130985
0.6161616161616162 0.005967362345474745
0.6262626262626263 0.004906497928501511
0.6363636363636365 0.0025195529903115593
0.6464646464646465 0.002917377146676574
0.6565656565656566 0.0023869449381899244
0.6666666666666667 0.0034478093551631867
0.6767676767676768 0.004508673772136524
0.686868686868687 0.003845633511528169
0.696969696969697 0.005436930136988162
0.7070707070707072 0.006099970397596407
0.7171717171717172 0.004376065720014861
0.7272727272727273 0.0010608644169732998
0.7373737373737375 0.0013260805212166101
0.7474747474747475 0.0035804174072848866
0.7575757575757577 0.0023869449381898984
0.7676767676767677 0.0007956483127299748
0.7777777777777778 0.001458688573338287
0.787878787878788 0.0011934724690949492
0.797979797979798 0.0026521610424332493
0.8080808080808082 0.0027847690945548816
0.8181818181818182 0.002784769094554912
0.8282828282828284 0.002917377146676542
0.8383838383838385 0.001989120781824937
0.8484848484848485 0.00013260805212166247
0.8585858585858587 0.000397824156364983
0.8686868686868687 0.0006630402606083123
0.8787878787878789 0.000795648312729966
0.888888888888889 0.00013260805212166247
0.8989898989898991 0.0001
0.9090909090909092 0.00013260805212166247
0.9191919191919192 0.0003978241563649874
0.9292929292929294 0.000132608052121661
0.9393939393939394 0.00013260805212166247
0.9494949494949496 0.000265216104243322
0.9595959595959597 0.0011934724690949622
0.9696969696969697 0.00026521610424332494
0.9797979797979799 0.0006630402606083051
0.98989898989899 0.00013260805212166247
1.0 0.00026521610424332494
1.0 0.0001
};
\addlegendentry{{edge hit rate 0.50, 2 epochs, $\sigma=0.2$}}
\addplot+ [no markers,opacity=0.5,blue,fill,draw,const plot mark right] table  {
x y
0.0 0.0001
0.010101010101010102 85.41142099921056
0.020202020202020204 2.4978324123153266
0.030303030303030304 1.6374579903011168
0.04040404040404041 1.3031780760121798
0.05050505050505051 1.0943926920040596
0.06060606060606061 0.895879102289388
0.07070707070707072 0.712773204014886
0.08080808080808081 0.6361813465659188
0.09090909090909091 0.5607059884966733
0.10101010101010102 0.4744005864441182
0.11111111111111112 0.4002650276305404
0.12121212121212122 0.3604059997744447
0.13131313131313133 0.3338333145370472
0.14141414141414144 0.29810533438592507
0.15151515151515152 0.23290177061012787
0.16161616161616163 0.23312507048607173
0.17171717171717174 0.21202323220931524
0.18181818181818182 0.18813014548325283
0.19191919191919193 0.17707680162399894
0.20202020202020204 0.15206721551821348
0.21212121212121213 0.1486060674410739
0.22222222222222224 0.11499943611142427
0.23232323232323235 0.11064508853050628
0.24242424242424243 0.0908830495094171
0.25252525252525254 0.08295590391338663
0.26262626262626265 0.07313070937182806
0.27272727272727276 0.06721326265929846
0.2828282828282829 0.051024021653321254
0.29292929292929293 0.048456073079959625
0.30303030303030304 0.034723130709371795
0.31313131313131315 0.03550468027517759
0.32323232323232326 0.03416488101951051
0.33333333333333337 0.02713093492725835
0.3434343434343435 0.02478628622984096
0.3535353535353536 0.024562986353896445
0.36363636363636365 0.020543588586895326
0.37373737373737376 0.016970790571783
0.38383838383838387 0.019427089207172645
0.393939393939394 0.015296041502199152
0.4040404040404041 0.01596594113003269
0.4141414141414142 0.011499943611142429
0.42424242424242425 0.010048494417503146
0.43434343434343436 0.010495094169392118
0.4444444444444445 0.011723243487086941
0.4545454545454546 0.007480545844141191
0.4646464646464647 0.010383444231419861
0.4747474747474748 0.009713544603586323
0.48484848484848486 0.00558249689861286
0.494949494949495 0.004689297394834776
0.5050505050505051 0.0035727980151122104
0.5151515151515152 0.0011164993797225596
0.5252525252525253 0.0013397992556670863
0.5353535353535354 0.0015630991316116008
0.5454545454545455 0.0023446486974173755
0.5555555555555556 0.0018980489455283722
0.5656565656565657 0.002567948573361887
0.5757575757575758 0.0008931995037780576
0.5858585858585859 0.0012281493176948292
0.595959595959596 0.002679598511334143
0.6060606060606061 0.0021213488214728866
0.6161616161616162 0.0018980489455283512
0.6262626262626263 0.0010048494417503146
0.6363636363636365 0.0005582496898612798
0.6464646464646465 0.0004465997518890288
0.6565656565656566 0.0010048494417503146
0.6666666666666667 0.0014514491936393277
0.6767676767676768 0.001116499379722572
0.686868686868687 0.0006698996278335358
0.696969696969697 0.0014514491936393435
0.7070707070707072 0.0013397992556670715
0.7171717171717172 0.0007815495658058004
0.7272727272727273 0.0032378482011954585
0.7373737373737375 0.0008931995037780477
0.7474747474747475 0.0014514491936393435
0.7575757575757577 0.0010048494417503038
0.7676767676767677 0.0015630991316116008
0.7777777777777778 0.0007815495658058004
0.787878787878788 0.0003349498139167679
0.797979797979798 0.0002232998759445144
0.8080808080808082 0.00022329987594451193
0.8181818181818182 0.0007815495658058004
0.8282828282828284 0.00022329987594451193
0.8383838383838385 0.00033494981391677157
0.8484848484848485 0.0004465997518890288
0.8585858585858587 0.00022329987594451193
0.8686868686868687 0.00033494981391677157
0.8787878787878789 0.0003349498139167679
0.888888888888889 0.00033494981391677157
0.8989898989898991 0.00022329987594451193
0.9090909090909092 0.0001
0.9191919191919192 0.0001116499379722572
0.9292929292929294 0.00044659975188902386
0.9393939393939394 0.0006698996278335431
0.9494949494949496 0.00044659975188902386
0.9595959595959597 0.0004465997518890288
0.9696969696969697 0.001116499379722572
0.9797979797979799 0.0001
0.98989898989899 0.0001
1.0 0.0001
1.0 0.0001
};
\addlegendentry{{edge hit rate 0.25, 2 epochs, $\sigma=0.2$}}
\addplot+ [no markers,opacity=0.5,gray!80,fill,draw,const plot mark right] table  {
x y
0.0 0.0001
0.010101010101010102 97.71655395196164
0.020202020202020204 0.24274162455061737
0.030303030303030304 0.1522680143802428
0.04040404040404041 0.11894649090814356
0.05050505050505051 0.1022341478664096
0.06060606060606061 0.08376804043140731
0.07070707070707072 0.06715886000104199
0.08080808080808081 0.062310217266711855
0.09090909090909091 0.05426353357995105
0.10101010101010102 0.04508206116813417
0.11111111111111112 0.04291564632939094
0.12121212121212122 0.029813994685562457
0.13131313131313133 0.034353149585786454
0.14141414141414144 0.024552701505757283
0.15151515151515152 0.02228312405564533
0.16161616161616163 0.0209420101078518
0.17171717171717174 0.02104517271922053
0.18181818181818182 0.01712499348720891
0.19191919191919193 0.015887042150784125
0.20202020202020204 0.013514302089303378
0.21212121212121213 0.013926952534778329
0.22222222222222224 0.009387797634554254
0.23232323232323235 0.011451049861928816
0.24242424242424243 0.009594122857291738
0.25252525252525254 0.008356171520866975
0.26262626262626265 0.00484864273433022
0.27272727272727276 0.0053644557911738605
0.2828282828282829 0.002475902672849474
0.29292929292929293 0.003920179232011689
0.30303030303030304 0.003817016620642939
0.31313131313131315 0.00433282967748658
0.32323232323232326 0.004951805345698948
0.33333333333333337 0.0026822278955869303
0.3434343434343435 0.0022695774501120176
0.3535353535353536 0.0020632522273745617
0.36363636363636365 0.0013411139477934725
0.37373737373737376 0.0013411139477934651
0.38383838383838387 0.001444276559162193
0.393939393939394 0.001444276559162193
0.4040404040404041 0.0008253008909498246
0.4141414141414142 0.001444276559162193
0.42424242424242425 0.0011347887250560153
0.43434343434343436 0.0007221382795810965
0.4444444444444445 0.0007221382795810965
0.4545454545454546 0.0005158130568436404
0.4646464646464647 0.0010316261136872808
0.4747474747474748 0.00030948783410618424
0.48484848484848486 0.000618975668212372
0.494949494949495 0.00030948783410618424
0.5050505050505051 0.0006189756682123685
0.5151515151515152 0.00020632522273745504
0.5252525252525253 0.0001
0.5353535353535354 0.000618975668212372
0.5454545454545455 0.0007221382795810926
0.5555555555555556 0.0001
0.5656565656565657 0.00020632522273745504
0.5757575757575758 0.0001
0.5858585858585859 0.000618975668212372
0.595959595959596 0.0009284635023185477
0.6060606060606061 0.000618975668212372
0.6161616161616162 0.0003094878341061825
0.6262626262626263 0.0007221382795811006
0.6363636363636365 0.0004126504454749101
0.6464646464646465 0.0007221382795811006
0.6565656565656566 0.00041265044547491463
0.6666666666666667 0.00010316261136872752
0.6767676767676768 0.00010316261136872866
0.686868686868687 0.0005158130568436376
0.696969696969697 0.00041265044547491463
0.7070707070707072 0.0003094878341061825
0.7171717171717172 0.00020632522273745732
0.7272727272727273 0.0001
0.7373737373737375 0.00010316261136872752
0.7474747474747475 0.00010316261136872866
0.7575757575757577 0.0001
0.7676767676767677 0.00020632522273745732
0.7777777777777778 0.0001
0.787878787878788 0.0001
0.797979797979798 0.0001
0.8080808080808082 0.0001
0.8181818181818182 0.00041265044547491463
0.8282828282828284 0.0001
0.8383838383838385 0.0001
0.8484848484848485 0.0001
0.8585858585858587 0.0001
0.8686868686868687 0.0001
0.8787878787878789 0.0001
0.888888888888889 0.0001
0.8989898989898991 0.0001
0.9090909090909092 0.0001
0.9191919191919192 0.0001
0.9292929292929294 0.0001
0.9393939393939394 0.0001
0.9494949494949496 0.0001
0.9595959595959597 0.0001
0.9696969696969697 0.0001
0.9797979797979799 0.0001
0.98989898989899 0.0001
1.0 0.0001
1.0 0.0001
};
\addlegendentry{{edge hit rate 0.02, 2 epochs, $\sigma=0.2$}}
\addplot+ [no markers,opacity=0.5,green,fill,draw,const plot mark right] table  {
x y
0.0 0.0001
0.010101010101010102 98.7481748153017
0.020202020202020204 0.2518251846982954
0.030303030303030304 0.0001
0.04040404040404041 0.0001
0.05050505050505051 0.0001
0.06060606060606061 0.0001
0.07070707070707072 0.0001
0.08080808080808081 0.0001
0.09090909090909091 0.0001
0.10101010101010102 0.0001
0.11111111111111112 0.0001
0.12121212121212122 0.0001
0.13131313131313133 0.0001
0.14141414141414144 0.0001
0.15151515151515152 0.0001
0.16161616161616163 0.0001
0.17171717171717174 0.0001
0.18181818181818182 0.0001
0.19191919191919193 0.0001
0.20202020202020204 0.0001
0.21212121212121213 0.0001
0.22222222222222224 0.0001
0.23232323232323235 0.0001
0.24242424242424243 0.0001
0.25252525252525254 0.0001
0.26262626262626265 0.0001
0.27272727272727276 0.0001
0.2828282828282829 0.0001
0.29292929292929293 0.0001
0.30303030303030304 0.0001
0.31313131313131315 0.0001
0.32323232323232326 0.0001
0.33333333333333337 0.0001
0.3434343434343435 0.0001
0.3535353535353536 0.0001
0.36363636363636365 0.0001
0.37373737373737376 0.0001
0.38383838383838387 0.0001
0.393939393939394 0.0001
0.4040404040404041 0.0001
0.4141414141414142 0.0001
0.42424242424242425 0.0001
0.43434343434343436 0.0001
0.4444444444444445 0.0001
0.4545454545454546 0.0001
0.4646464646464647 0.0001
0.4747474747474748 0.0001
0.48484848484848486 0.0001
0.494949494949495 0.0001
0.5050505050505051 0.0001
0.5151515151515152 0.0001
0.5252525252525253 0.0001
0.5353535353535354 0.0001
0.5454545454545455 0.0001
0.5555555555555556 0.0001
0.5656565656565657 0.0001
0.5757575757575758 0.0001
0.5858585858585859 0.0001
0.595959595959596 0.0001
0.6060606060606061 0.0001
0.6161616161616162 0.0001
0.6262626262626263 0.0001
0.6363636363636365 0.0001
0.6464646464646465 0.0001
0.6565656565656566 0.0001
0.6666666666666667 0.0001
0.6767676767676768 0.0001
0.686868686868687 0.0001
0.696969696969697 0.0001
0.7070707070707072 0.0001
0.7171717171717172 0.0001
0.7272727272727273 0.0001
0.7373737373737375 0.0001
0.7474747474747475 0.0001
0.7575757575757577 0.0001
0.7676767676767677 0.0001
0.7777777777777778 0.0001
0.787878787878788 0.0001
0.797979797979798 0.0001
0.8080808080808082 0.0001
0.8181818181818182 0.0001
0.8282828282828284 0.0001
0.8383838383838385 0.0001
0.8484848484848485 0.0001
0.8585858585858587 0.0001
0.8686868686868687 0.0001
0.8787878787878789 0.0001
0.888888888888889 0.0001
0.8989898989898991 0.0001
0.9090909090909092 0.0001
0.9191919191919192 0.0001
0.9292929292929294 0.0001
0.9393939393939394 0.0001
0.9494949494949496 0.0001
0.9595959595959597 0.0001
0.9696969696969697 0.0001
0.9797979797979799 0.0001
0.98989898989899 0.0001
1.0 0.0001
1.0 0.0001
};
\addlegendentry{{LSST 30 epochs}}
\addplot+ [no markers,opacity=0.5,black,fill,draw,const plot mark right] table  {
x y
0.0 0.0001
0.010101010101010102 98.99999999999999
0.020202020202020204 0.0001
0.030303030303030304 0.0001
0.04040404040404041 0.0001
0.05050505050505051 0.0001
0.06060606060606061 0.0001
0.07070707070707072 0.0001
0.08080808080808081 0.0001
0.09090909090909091 0.0001
0.10101010101010102 0.0001
0.11111111111111112 0.0001
0.12121212121212122 0.0001
0.13131313131313133 0.0001
0.14141414141414144 0.0001
0.15151515151515152 0.0001
0.16161616161616163 0.0001
0.17171717171717174 0.0001
0.18181818181818182 0.0001
0.19191919191919193 0.0001
0.20202020202020204 0.0001
0.21212121212121213 0.0001
0.22222222222222224 0.0001
0.23232323232323235 0.0001
0.24242424242424243 0.0001
0.25252525252525254 0.0001
0.26262626262626265 0.0001
0.27272727272727276 0.0001
0.2828282828282829 0.0001
0.29292929292929293 0.0001
0.30303030303030304 0.0001
0.31313131313131315 0.0001
0.32323232323232326 0.0001
0.33333333333333337 0.0001
0.3434343434343435 0.0001
0.3535353535353536 0.0001
0.36363636363636365 0.0001
0.37373737373737376 0.0001
0.38383838383838387 0.0001
0.393939393939394 0.0001
0.4040404040404041 0.0001
0.4141414141414142 0.0001
0.42424242424242425 0.0001
0.43434343434343436 0.0001
0.4444444444444445 0.0001
0.4545454545454546 0.0001
0.4646464646464647 0.0001
0.4747474747474748 0.0001
0.48484848484848486 0.0001
0.494949494949495 0.0001
0.5050505050505051 0.0001
0.5151515151515152 0.0001
0.5252525252525253 0.0001
0.5353535353535354 0.0001
0.5454545454545455 0.0001
0.5555555555555556 0.0001
0.5656565656565657 0.0001
0.5757575757575758 0.0001
0.5858585858585859 0.0001
0.595959595959596 0.0001
0.6060606060606061 0.0001
0.6161616161616162 0.0001
0.6262626262626263 0.0001
0.6363636363636365 0.0001
0.6464646464646465 0.0001
0.6565656565656566 0.0001
0.6666666666666667 0.0001
0.6767676767676768 0.0001
0.686868686868687 0.0001
0.696969696969697 0.0001
0.7070707070707072 0.0001
0.7171717171717172 0.0001
0.7272727272727273 0.0001
0.7373737373737375 0.0001
0.7474747474747475 0.0001
0.7575757575757577 0.0001
0.7676767676767677 0.0001
0.7777777777777778 0.0001
0.787878787878788 0.0001
0.797979797979798 0.0001
0.8080808080808082 0.0001
0.8181818181818182 0.0001
0.8282828282828284 0.0001
0.8383838383838385 0.0001
0.8484848484848485 0.0001
0.8585858585858587 0.0001
0.8686868686868687 0.0001
0.8787878787878789 0.0001
0.888888888888889 0.0001
0.8989898989898991 0.0001
0.9090909090909092 0.0001
0.9191919191919192 0.0001
0.9292929292929294 0.0001
0.9393939393939394 0.0001
0.9494949494949496 0.0001
0.9595959595959597 0.0001
0.9696969696969697 0.0001
0.9797979797979799 0.0001
0.98989898989899 0.0001
1.0 0.0001
1.0 0.0001
};
\addlegendentry{{LSST 460 epochs}}
\end{axis}
\end{tikzpicture}